\documentclass{emulateapj}
\usepackage{natbib}
\usepackage{graphicx}
\usepackage{amsmath}
\usepackage[version=3]{mhchem}
\usepackage{epsfig}
\begin{document}
\title{Multilayer formation and evaporation of deuterated ices in
  prestellar and protostellar cores}
\author{Vianney Taquet$^{1,2}$,  Steven B. Charnley$^{2}$, and Olli
  Sipil{\"a}$^3$}
\altaffiltext{1}{NASA Postdoctoral Program Fellow} 
\altaffiltext{2}{Astrochemistry Laboratory and The
  Goddard Center for Astrobiology, Mailstop 691, NASA Goddard Space
  Flight Center, 8800 Greenbelt Road, Greenbelt, MD 20770, USA}
\altaffiltext{3}{Department of Physics, PO Box 64, 00014 University of
  Helsinki, Finland} 
    \date{Received - ; accepted -}
\begin{abstract}

Extremely large deuteration of several molecules has been observed
towards prestellar cores and low-mass protostars for a decade. 
New observations performed towards low-mass protostars suggest that
water presents a lower deuteration in the warm inner gas than in the
cold external envelope.  
%
We coupled a gas-grain astrochemical model with a one-dimension model
of collapsing core to properly follow the formation and the deuteration of
interstellar ices as well as their subsequent evaporation in the low-mass
protostellar envelopes with the aim of interpreting the spatial and
temporal evolutions of their deuteration. 
{The astrochemical model follows the formation and the evaporation of
ices with a multilayer approach and also includes a state-of-the-art
deuterated chemical network by taking the spin states of H$_2$ and
light ions into account. }
%
Because of their slow formation, interstellar ices are chemically
heterogeneous and show an increase of their deuterium fractionation
towards the surface.  
The differentiation of the deuteration in ices induces an evolution of
the deuteration within protostellar envelopes. 
The warm inner region is poorly deuterated because it includes the
whole molecular content of ices while the deuteration predicted in the
cold external envelope scales with the highly deuterated surface of
ices. 
We are able to reproduce the observed evolution of water deuteration
within protostellar envelopes but we are still unable to predict the
super-high deuteration observed for formaldehyde and methanol.  
{Finally, the extension of this study to the deuteration of complex
organics (COMs), important for the prebiotic chemistry, shows a good
agreement with the observations, suggesting that we can use the
deuteration to retrace their mechanisms and their moments of formation. }

 \end{abstract}



\section{Introduction}

The early stages of low-mass star formation are characterized by extremely
enhanced molecular deuteration observed for various gaseous species.
In spite of the low cosmic elemental reservoir of deuterium
\citep[$1.5 \times 10^{-5}$,][]{Linsky2003}, the deuterium
fractionation{\footnote 
{In the following of this manuscript, the deuterium fractionation,
  also referred as the D/H abundance ratio, of a deuterated species XD
  stands for the abundance ratio between this deuterated species and
  its main isotopologue XH: D/H(XD) = [XD]/[XH].}} of simply
and doubly deuterated species can reach $\sim 50$ \% and $\sim 15$ \%,
respectively in low-mass prestellar cores and Class 0 protostars
\citep{Parise2006, Bergman2011} while triply deuterated 
isotopologues have also been detected \citep{Parise2004, Roueff2005}.   

The deuterium chemistry in the gas phase is believed to be triggered
by the deuterium exchange between HD and H$_{3}^+$, as the deuterium
reservoir is likely initially located in HD, via the reaction 
\begin{equation} 
\textrm{H}_3^+ + \textrm{HD} \rightarrow
\textrm{H}_2\textrm{D}^+ + \textrm{H}_2 + 232~\textrm{K}. \label{deut_reac} 
\end{equation}  
which is in competition with the reaction between  H$_{3}^+$ and CO. 
In dense and cold prestellar cores, the CO freeze-out at the surface of
interstellar grains increases the reactivity of reaction
(\ref{deut_reac}) as well as its deuterated analogues through the
decrease of the destruction of H$^+_3$ via the
reaction between  H$_{3}^+$ and CO.  
For high densities and high CO depletions, the gas
phase D/H abundance ratio of deuterated species can easily reach
100 \% \citep{Roberts2003}.   
Since reaction (\ref{deut_reac}) is exothermic, its backward process
cannot occur in the cold conditions ($T < 20$ K) found in prestellar cores
if only the para spin state of H$_2$ is considered. 
However, due to its higher internal energy, the ortho spin state
allows the reaction between H$_2$ and H$_2$D$^+$ to occur, even at low
temperatures. Therefore, the ortho/para ratio of H$_2$ opr(H$_2$)
strongly influences the deuteration of gaseous species in prestellar cores
\citep{Flower2006}.  

Millimetric observations performed towards Class 0 protostars have
shown that various molecules, mostly formed at the surface of
interstellar grains in molecular clouds and evaporated in the warm
inner regions of protostellar envelopes at $T>100$ K (the so-called
hot corinos), show different deuterations. 
For example, in the low-mass protostar prototype IRAS 16293-2422
(hereafter I16293), formaldehyde and methanol show higher deuterations
\citep[with D/H ratios of 15 and 40 \%, respectively,
see][]{Loinard2001, Parise2002, Parise2004} than ammonia and water
\citep[$\sim 10$ and 3 \%, respectively, see][]{vanDishoeck1995, Coutens2012}.  
The deuterium fractionation can be used as a chemical tool to retrace
the history of interstellar ices and gas-grain astrochemical models
have been developed to interpret the observations of low-mass protostars
\citep{Cazaux2011, Taquet2012b, Taquet2013a}.   
The lower deuterium fractionation of water with respect to that of
formaldehyde and methanol is likely due to a different epoch of formation. 
Water would be mainly produced for typical molecular cloud conditions
(low density and warm temperature) while most of formaldehyde and
methanol would be formed afterwards, during the colder and denser
prestellar phase.  
{A low opr(H$_2$), lower than $10^{-3}$, is also required to reproduce
the observed deuterations \citep{Taquet2013a}. }

Because of its high sensitivity to the variation of physical and
chemical parameters, the deuterium fractionation can also be used to probe
the chemical evolution through the formation of low-mass stars like the
Sun. 
The comparison of the molecular content observed in prestellar cores and
low-mass protostars with that found in Solar System comets and
meteorites shows that the deuterium fractionation of water and
organics decreases by several orders of magnitude \citep[from 1 - 100 \% to
0.01 - 0.1 \%, see][]{Ceccarelli2014}. 
{The decrease of the D/H ratio with the evolutionary stage of star
formation can be interpreted by the chemical alteration of the molecular
content in protoplanetary disks and in the solar nebula. Molecules
observed in protostellar envelopes will be eventually recycled in
protoplanetary disks due to the high turbulent mixing and the strong
UV and X-ray radiation fields irradiating the disks 
\citep{Furuya2013, Albertsson2014}. } 
The difference of deuteration could also come from the different scales
probed by the astronomical observations and by the study of solar system
bodies. {Most of the meteorites and comets studied in our Solar System
were initially relatively closed to the Sun ($< 50$ AU) while most of
sub-millimetric astronomical observations of low-mass protostars are
sensitive to much larger scales of several hundreds/thousands of AU
\citep[see][]{Ceccarelli2014}.  }

New observations of deuterated water performed with the {\it Herschel Space
Observatory} \citep{Coutens2013a, Coutens2013b} as well
as ground-based interferometers \citep{Persson2013, Visser2013,  Taquet2013b,
  Persson2014} towards low-mass protostars have allowed astronomers to
probe the deuteration of warm water originating from the inner regions
of protostellar envelopes ($< 100$ AU), and suggest that warm water
shows a lower deuteration ([HDO]/[H$_2$O] $\sim 0.1 - 1$ \%) than the
values derived in previous studies with single-dish telescopes
\citep[1 - 10 \%,][]{Liu2011, Coutens2012}.  
Moreover, \citet{Coutens2013a, Coutens2013b} showed that the
deuteration of cold water originating from the external envelope would
be higher ($\sim 1$ \%) than the deuteration of warm inner water.
These observations, therefore, suggest an evolution of the deuteration
within protostellar envelopes encompassed by the large beam of
single-dish telescopes used so far to study the deuteration of most of molecules. 
Consequently, astrochemical models have to to be coupled with evolutionary
models to properly follow the deuteration within prestellar cores and
protostellar envelopes. 
A previous attempt has been made by \citet{Aikawa2012} who followed
the evolution of the molecular deuteration from molecular clouds to
protostellar envelopes through the coupling of a gas-grain
astrochemical model with a one-dimensional radiative hydrodynamic
model. 
These authors showed an evolution of the chemical abundances of icy species and
complex organics with the age of the protostar and gave predictions
for their deuterium fractionation.
However, they used a standard two-phase gas-grain astrochemical model
which did not allow them to distinguish the chemical processes between
the ice surface and the ice mantle, a crucial aspect for the modelling
of deuterated ices \citep{Taquet2012b}, and did not include the effect of the spin
states of H$_2$ on the deuterium chemistry.

The ortho/para ratio of H$_2$ is believed to evolve in molecular
clouds from a high ortho/para ratio, induced by its
formation at the surface of interstellar grains as recently observed in
laboratory experiments \citep{Watanabe2010} to its low Boltzmann value
($< 10^{-4}$), with a timescale comparable to the lifetime of molecular
clouds \citep[$\sim 10^7$ yr, see][]{Flower2006}. 
The slow decrease of the opr(H$_2$) is supported by the non-detection
of DCO$^+$ towards interstellar clouds outside dense cores which has
been attributed by \citet{Pagani2011} to an opr(H$_2$) higher to 0.1.
Since interstellar ices start to form in translucent regions of molecular
clouds, when the visual extinction exceeds $\sim 3$ mag
\citep{Whittet1988}, a part of these interstellar ices would likely
form when the opr(H$_2$) is still high.

In this work, we aim at self-consistently studying the deuterium
chemistry from molecular clouds to protostellar envelopes. 
For this purpose, we coupled our astrochemical model GRAINOBLE with a
simple, but accurate, one-dimensional evolutionary model of core
collapse and we included the spin states of H$_2$ and light ions that
are of crucial importance for the deuterium chemistry. We will focus
our study on the formation, the deuteration and the evaporation of
interstellar ices and complex organics, {which are believed to be
relevant for the prebiotic chemistry found in the Solar System
\citep[see][]{Pizzarello2005, Mumma2011}}.
The physical code and the astrochemical model are
described in section 2. In section 3, we present the evolution of the
molecular deuteration by focusing  our study on the main ice components
and on complex organics formed at the surface of grains. We then
discuss the implications of our work in section 4 and present our
conclusions in section 5.


\section{Model}

\subsection{Dynamical Model}

This section describes the physical model used in this work to follow
the formation of a prestellar core and its collapse, giving birth to
a central protostar surrounded by a protostellar envelope. 
The physical evolution can be divided into three steps: 
1) the static formation of the dense core for $t < 0$, where $t= 0$ is
defined as the beginning of the core contraction; 
2) the collapse of the dense core for $0 < t <
t_{\textrm{FF}}$, where $t_{\textrm{FF}}$ is the free-fall timescale
of the core;
3) the birth of the protostar and the collapse of the protostellar
envelope for $t_{\textrm{FF}} < t < t_{\textrm{FF}} +
t_{\textrm{C0}}$, where $t_{\textrm{C0}}$ is the timescale of the
Class 0 stage, defined as the moment when the mass of the central
protostar reaches half of the total (protostar+envelope) mass.

We considered 182 shells of gas and dust initially spaced between $1
\leq r \leq 3 \times 10^4$ AU so they are spaced at logarithmic
intervals for specific times during the Class 0 stage. 
Because the core is evolving dynamically, the position of each shell
evolves with time, so it is necessary to solve the chemistry in a
frame of reference comoving with the gas in the Lagrangian coordinates.

\subsubsection{Static formation of the core}

We followed the formation of the dense core by assuming a static
contraction, starting from a homogeneous translucent 
sphere of a density $n_{\textrm{H,ini}}$ embedded in surroundings by
a visual extinction $A_{V\textrm{,ext}}$.
During its static contraction, the core keeps a Plummer-like density profile
obtained from this equation
\begin{equation}
n_{\textrm{H}} = \frac{n_{\textrm{H,0}}}{(1+(r/R_{\textrm{f}})^2)^{\eta/2}}
\end{equation}
where $n_{\textrm{H,0}}$ is the central density, $R_{\textrm{f}}$ is
the characteristic radius inside which the density is
uniform. Outside $R_{\textrm{f}}$, the density follows a power-law
r$^{-\eta}$ when $r >> R_{\textrm{f}}$. 
$n_{\textrm{H,0}}$ increases with time between $n_{\textrm{H,ini}}$
and the central density of the observed core $n_{\textrm{H,0,obs}}$
{(see section \ref{free_parameters})} to reach different selected
  intermediate central densities 
$n_{\textrm{H,0}}$ in timescales given  by their respective free-fall
timescale. The choice of the intermediate {and final} central 
densities (and of their timescale) is explained in section
\ref{free_parameters} and is based on the observational estimates of
molecular clouds ages.  
As $R_f$ is given by the product of the sound speed and the free-fall
time of the central density $n_{\textrm{H,0}}$ \citep{Keto2010},
$R_{\textrm{f}}$ decreases with $1/\sqrt{n_{\textrm{H,0}}}$.

\subsubsection{Free-fall collapse}

After its static formation, the dense core undergoes a free-fall
collapse following the model developed by \citet{Whitworth2001}.
At $t = 0$, the collapse is triggered artificially, the density and velocity
profiles as well as the accretion rate are followed with time from
equations (4), (6), and (7) of \citet{Whitworth2001}. 

Due to the rapid accretion of the inner shells towards the center, a
protostar is formed at the center of the envelope once $t =
t_{\textrm{FF}}$, with  
\begin{equation}
t_{\textrm{FF}}  = \sqrt{\frac{3 \pi}{32 G n_{\textrm{H,0,obs}} m_p}}
\end{equation}
where $m_p$ is the proton mass. The gravitational energy of the
collapsed mass is released radiatively as the material falls onto the
protostar radius $R_{\star}$.
The luminosity of the protostar is, therefore, given by 
\begin{equation}
L_{\star} = \frac{G M_{\star} \dot{M}_{acc}}{R_{\star}}
\end{equation}
where $M_{\star}$ is the mass of the
protostar, and $\dot{M}_{acc}$ is the accretion rate. 
$\dot{M}_{acc}$ and $M_{\star}$ are directly given the dynamical
model and we derived $R_{\star}$ from the work by \citet{Hosokawa2009}
who computed the evolution of the star radius as function of the
stellar mass, for different constant accretion rates. Since the
accretion rate computed by our code evolves with time, we extrapolated their
Fig. 14 by computing the averaged accretion rate at each timestep
relative to the beginning of the protostar formation in order to derive an
estimate of $R_{\star}$ for each timestep.

\subsubsection{Physical parameters} \label{free_parameters}

The density profile reached at the end of the static formation
phase aims at modelling the starless core L1498. 
L1498 is thought to have a central density $n_{\textrm{H,0,obs}} = 2
\times 10^5$ cm$^{-3}$ and tends to be extended with a radius
$R_{\textrm{f}} = 1 \times 10^4$ AU and presents a steep slope ($\eta = 4$)
\citep{Tafalla2002}.

\begin{table*}[htp]
\centering
\caption{Physical models and their parameters considered in this work
  (see text for a description of the timescales).}
\begin{tabular}{l c c c c c c c c}
\hline
\hline
Name & $n_{\textrm{H,0,obs}}$ & $R_{f}$ & $\eta$ & $M$ & $t_{\textrm{MC}}$ & $t_{\textrm{SC}}$ &
$t_{\textrm{CC}}$ & $t_{\textrm{C0}}$ \\
 & (cm$^{-3}$) & (AU) & & ($M_{\odot}$) & (yr) & (yr) & (yr) & (yr) \\
\hline
A-F & $2 \times 10^5$ & $1.1 \times 10^4$ & 4 & 5.0 & $1.7 \times 10^6$
& $2.3 \times 10^5$ & $9.8 \times 10^4$ & $1.0 \times 10^5$ \\
A-S & $2 \times 10^5$ & $1.1 \times 10^4$ & 4 & 5.0 & $3.4 \times 10^6$
& $4.6 \times 10^5$ & $9.8 \times 10^4$ & $1.0 \times 10^5$  \\
\hline
Observations & & & & & $3 - 6 \times 10^6$ & $2 - 7 \times 10^5$ & $2 - 16
\times 10^4$ & $0.4 - 1.6 \times 10^5$ \\
\hline
\end{tabular}
\tablecomments{{
 $t_{\textrm{MC}}$: lifetime of molecular clouds. 
 $t_{\textrm{SC}}$: lifetime of static cores. 
 $t_{\textrm{CC}}$: lifetime of contracting cores. 
 $t_{\textrm{C0}}$: lifetime of Class 0 protostars.
}}
\label{core_param}
\end{table*}

The timescale for the static formation needed to reach the observed
profile of the core is based on the timescale estimates derived
from  observations and models of molecular clouds.  
The lifetime of molecular clouds is still controversial as two main
contradictory theories have been developed. The first one is a quasi-static
theory in which clouds are closed to equilibrium, giving long
timescales for the evolution of molecular clouds of $\sim 10$ Myr
\citep{Mouschovias2006}. A more recent view of cloud evolution has
been proposed, which is based on the importance of supersonic
turbulent flows leading to a lack of equilibrium and giving,
therefore, shorter timescales of 3-5 Myr \citep{Hartmann2001}. 
Chemistry can be used as a tracer for the age of molecular
clouds. From the non-detection of DCO$^+$ in dark clouds,
\citet{Pagani2011} estimated an upper limit for the clouds age of 6
Myr. 
Consequently, we can reasonably assess that molecular clouds have a total
lifetime of a few Myr.
We then derived the total timescale of static formation based on the
molecular cloud lifetime estimate and on the free-fall timescale of
seven different Plummer-like cores with selected central densities
between $n_{\textrm{H,ini}}$ and $n_{\textrm{H,0,obs}}$
\citep[see also][for a similar method of static formation]{Lee2004a}.  
Table \ref{core_param} summarizes the physical properties of the core
and compares the different timescales $t_{\textrm{MC}}$ spent in the
molecular cloud phase (i.e, when the central density is lower than $5
\times 10^4$ yr).  
{Model A-F is a model with a standard static formation while model
A-S is a model with a two times longer static formation timescale. }

\subsection{Radiative transfer}

We used the radiative transfer code DUSTY \citep{Ivezic1997} to
compute the dust temperature profile of the prestellar and
protostellar cores at each timestep. 
We assumed $T_{\textrm{g}} = T_{\textrm{d}}$, a reasonable assumption for
protostellar envelopes \citep[see for instance][]{Maret2004} and for
the central region of dark cores but not necessarily true at the outer
edges of dark cores where the gas and dust are thermally decoupled due
to the lower density \citep[see][]{Galli2002, Sipila2012}. 
We verified that this assumption is not important for the chemical
abundance and deuteration of ices by running two
simulations for the streamline at $r = 2 \times 10^4$ AU with two
gas phase {temperatures: $T_g = T_d = 16$ K} (the temperature assumed in
the reference model) and $T_g = 10$ K (corresponding to the kinetic
temperature derived by \citet{Tafalla2004} at the edge of L1498). 
The largest differences for the deuteration and the chemical
composition of ices are found to be about 30 \%.  
The opacities of dust grains are taken from column 5 in Table 5 of
\citet{Ossenkopf1994} corresponding to grains showing a MRN size
distribution covered by thin ice mantles at $n_{\textrm{H}} = 10^6$ cm$^{-3}$. This
data has been successfully used in many works, such as those by
\citet{vanderTak2000}  and \citet{Evans2001} to model the emission of
prestellar and protostellar envelopes.

The thermal structure of the prestellar core is derived from a slab
geometry in which the core is irradiated by the Interstellar Radiation
Field (ISRF) with a spectrum taken from \citet{Evans2001}. We fixed
the temperature at the edge of the cores to 17 K, following the values
derived by \citet{WardThompson2002} towards L1498 at $3
\times 10^4$ AU ($\sim 3-4$ \arcmin).
Once the protostar has formed at the center of the envelope, the
temperature profile is computed with the sphere geometry
structure. The effective temperature of the protostar and the grain
temperature at $r_{\textrm{min}} = 1$ AU, derived from the protostar
luminosity, are given as input parameters to the DUSTY code.

\subsection{Astrochemical model}
 
\subsubsection{Description of the model} 

To follow the interstellar gas-grain chemistry, we used our
astrochemical model GRAINOBLE presented in 
previous studies \citep{Taquet2012a, Taquet2012b,
  Taquet2013a}. GRAINOBLE couples the gas-phase and grain-surface
chemistry with the rate equations approach introduced by
\citet{Hasegawa1992}. The grain surface chemistry processes are the
following: 

i) The accretion of gas phase species onto the surface of spherical grains
with a constant grain radius $r_d = 0.1$ $\mu$m. \\ 
ii) The diffusion of adsorbed species via thermal hopping. \\
iii) The reaction between two particles once they meet in the same
site via the Langmuir-Hinshelwood mechanism. The reaction rate is given
by the product of the number of times that the two reactants meet each other
and the transmission probability $P_r$ of reaction. \\
iv) The desorption of adsorbed species into the gas phase, via several processes: \\
- thermal desorption, which exponentially depends on the
binding energy of each species $E_b$ relative to the substrate (see Table
\ref{listEb} for a list of binding energies for selected species). \\ 
- cosmic-ray induced heating of grains following \citet{Hasegawa1993a}
and adapted for the binding energies considered in this work. \\ 
- chemical desorption caused by the energy release of exothermic
reactions. Following \citet{Garrod2007}, we assumed a value of 0.012
for the factor $a$ (the ratio of the surface-molecule bond frequency
to the frequency at which energy is lost to the grain surface) for all
surface reactions as it seems to be the most consistent value given by
molecular dynamics simulations \citep{Kroes2005}.  \\
- UV photodissociation and photodesorption. Photodissociation and
photodesorption rates of hydrogenated species are derived from molecular
dynamics simulations \citep{Andersson2008} while photodesorption rates
  of CO, N$_2$, O$_2$ are taken from the experimental studies by
  \citet{Fayolle2011, Fayolle2013}.  

\begin{table}[htp]
\centering
\caption{Selected surface species and their binding energies.}
\begin{tabular}{l c c c c}
\hline
\hline
Species & E$_{b,\textrm{bare}}$ & E$_{b,\textrm{wat}}$ & E$_{b,\textrm{pure}}$  & Ref. \\
 & (K) & (K) & (K) &  \\
\hline
H & 660 & 500 & 45 & 1, 2,  3 \\
C & 800 & 800 &   800 & 4 \\
O & 800 & 800 &   800 & 4  \\
N & 800 & 800 &  800 & 4  \\
S & 1100 & 1100 & 1100 & 4  \\
H$_2$ & 640 & 500    &    45 & 1,  2,  3 \\
O$_2$ & 895 & 915 & 915 & 5,  6 \\
N$_2$ & 790 & 790 & 790 & 7 \\
CO & 830 & 1150 & 855 & 5,  8,  6 \\
CO$_2$ & 2270 & 2690 & 2270 & 5,  9 \\
H$_2$O & 1870 & 5775 & 5775 & 10,  11 \\
NH$_3$ & 5535 & 5535 & 3080 & 8,  12  \\
CH$_4$ & 1090  &  1090 & 1090 & 13  \\
HCN & 2100 & 2100 & 2100 & 14  \\
H$_2$CO & 3260 & 3260 & 3765 & 15  \\
CH$_3$OH & 5530 & 5530 & 4230 & 8,  12 \\
C$_2$H$_2$ & 2600 & 2600 & 1830 & 16  \\
C$_2$H$_4$ & 3600 & 3600 & 2060 & 17	 \\
HCOOH & 5570 & 5570 & 5000 & 8,  18 \\
CH$_3$CN & 4680 & 4680 & 4680 & 8 \\
HCOOCH$_3$ & 4630 & 4630 & 4000 & 19,  18 \\
CH$_3$CHO & 3800 & 3800 & 3800 & 18     \\
CH$_3$OCH$_3$ & 4230 & 4230 & 3300 & 19,  18  \\
CH$_3$CH$_2$OH & 6795 & 6795 & 5200 & 19,  18  \\
NH$_2$OH & 6250 & 6250 & 6250 & 20  \\
CH$_3$COOH & 8155 & 8155 & 6300 & 19,  18 \\
CN & 1600 & 1600 & 1600 & 21 \\
CH & 870 & 870 & 870 & 22 \\
NH & 2380 & 2380 & 1560 & 23\\
CH$_2$ & 945 & 945 & 945 & 24 \\
NH$_2$ & 3960 & 3960 & 2320 & 25 \\
CH$_3$ & 1017 & 1017 & 1017 & 26 \\
OH & 1300  &  2820 & 2820 & 27,  11 \\
HCO & 2050 & 2200 & 2300 & 28 \\
CH$_3$O & 3800 & 3800 & 4000 & 29  \\
C$_2$ & 1600 & 1600 & 1600 & 30 \\
C$_2$H & 2100 & 2100 & 1715 & 31 \\
\hline
\end{tabular}
\label{listEb}
\tablecomments{
$^1$: \citet{Katz1999}; 
$^2$: \citet{Hornekaer2005};
$^3$: \citet{Vidali1991};
$^4$: \citet{Tielens1987};
$^5$: \citet{Noble2012a};
$^6$: \citet{Acharyya2007};
$^7$: \citet{Bisschop2006};
$^8$: \citet{Collings2004}; 
$^{9}$: \citet{Sandford1990}; 
$^{10}$: \citet{Avgul1970}; 
$^{11}$: \citet{Fraser2001}; 
$^{12}$: \citet{Sandford1993}; 
$^{13}$: \citet{Herrero2010}; 
$^{14}$: $E_b$(C)+$E_b$(N)+$E_b$(H); 
$^{15}$: \citet{Noble2012b}; 
$^{16}$: $E_b$(C$_2$)+2$\times$$E_b$(H); 
$^{17}$: $E_b$(C$_2$)+4$\times$$E_b$(H); 
$^{18}$: \citet{Oberg2009}; 
$^{19}$: \citet{Lattelais2011}; 
$^{20}$ \citet{Congiu2012}; 
$^{21}$: $E_b$(N)+$E_b$(C); 
$^{22}$: $E_b$(C)+($E_b$(CH$_4$)-$E_b$(C))/4; 
$^{23}$: $E_b$(N)+($E_b$(NH$_3$)-$E_b$(N))/3; 
$^{24}$: $E_b$(C)+($E_b$(CH$_4$)-$E_b$(C))/2; 
$^{25}$: $E_b$(N)+2$\times$($E_b$(NH$_3$)-$E_b$(N))/3; 
$^{26}$: $E_b$(C)+3$\times$($E_b$(CH$_4$)-$E_b$(C))/4; 
$^{27}$: \citet{Cuppen2007};
$^{28}$: ($E_b$(CO)+$E_b$(H$_2$CO))/2;
$^{29}$:  ($E_b$(H$_2$CO)+$E_b$(CH$_3$OH))/2;
$^{30}$:  2$\times$$E_b$(C);
$^{31}$:  2$\times$$E_b$(C)+$E_b$(H);
} 
\end{table}

In this work, we adopted the multilayer approach developed by
\citet{Hasegawa1993b} \citep[see also][for other models adopting a
multilayer approach]{Charnley2009, Garrod2011, Garrod2013,
  Vasyunin2013} to follow the multilayer formation and
evaporation of interstellar ices since it allows a better treatment of the
ice evaporation than the approach used in \citet{Taquet2012a,
  Taquet2012b, Taquet2013a}. 
The multilayer approach proposed by \citet{Hasegawa1993b} considers
three sets of differential equations: one for gas phase species, one
for surface species, and one for bulk species. The equations governing
chemical abundances on the surface and in the bulk are linked via an
additional term, proportional to the rate of growth or evaporation of the
grain mantle. Surface species are continuously trapped into the bulk
because of the accretion of new particles, {unlike the approach
by \citet{Taquet2012a} in which the trapping is carried out
one layer at a time. }
Using the chemical network and the reference physical conditions
adopted in \citet{Taquet2012a}, we verified that the two methods give
similar ice chemical compositions. The largest difference is found for
methanol which shows a difference of 50 \% at $t=10^6$ yr. 

\subsubsection{Gas phase chemical network}

The initial chemical network was taken from the KIDA database and includes 11
elements (H, He, C,  N, O, Si, S, Fe, Na, Mg, Cl) {and 6250
  reactions}. It has then been extended to include the spin states of
H$_2$, H$_2^+$, and H$_3^+$ and the deuterated isotopologues of
hydrogenated species with four or less atoms, with some exceptions
(see below). 

We added the spin states of H$_2$, H$_2^+$, and H$_3^+$ into the KIDA
chemical network following the method summarized by
\citet{Sipila2013} in which the separation of each chemical reaction including
H$_2$, H$_2^+$, or H$_3^+$ depends on the type of the reaction. 
{There are about 1150 reactions containing H$_2$, H$_2^+$, and
  H$_3^+$, the ortho/para separation process results in $\sim$2150
  reactions, increasing the number of reactions from 6250 to 7150 reactions.}

The chemical network including the spin states of H$_2$, H$_2^+$,
H$_3^+$ was then deuterated with the routine developed by
\citet{Sipila2013} \citep[see also][for other generated deuterated
networks]{Rodgers1996, Aikawa2012, Albertsson2013}. 
Since we aim at following the deuteration of interstellar ices, we are
particularly interested in the evolution of the atomic [D]/[H]
abundance ratio in the gas phase. 
We, therefore, included deuterated forms of species with up to four
atoms. All reactions including an hydrogenated species are cloned by
assuming same rate coefficients. When the final position of the D atom
is ambiguous, giving several sets of possible products, statistical
branching ratios have been assumed. 
{2840 reactions containing species with up to four atoms have been
  deuterated, resulting in new 7900 reactions added to the network.}
We also deuterated the reactions involved in the gas phase chemical
network of H$_2$O, NH$_3$, H$_2$CO, and CH$_3$OH and which contain
molecules with more than four atoms \citep[see][for a 
schematic picture of the water, ammonia, and formaldehyde
networks]{Roberts2000}. 

{Although we aim at investigating the deuterium fractionation of
  specific complex organics in the hot corino, we chose not to
  deuterate their formation and destruction reactions in the gas
  phase. Following recent state-of-the-art gas-grain astrochemical
  models, the current scenario is that complex organics are mostly
  formed at the surface of interstellar grains \citep{Garrod2008}
  while gas phase reactions would alter their abundances and
  deuterations in a timescale higher than $10^4$ yr
  \citep{Charnley1997, Aikawa2012}. 
  Due to the free-fall collapse model used in this work, molecules
  formed at the surface of grains stay in the warm gas phase after
  their evaporation for $\sim$100 yr before being accreted by 
  the protostar. Gas phase reactions are therefore considered as negligible in
  the formation and the deuteration of complex organics. }

The atomic [D]/[H] abundance ratio and the ortho/para ratio of H$_2$
are mostly governed by deuterium fractionation reactions involving
H$^+$, H$_2$, H$_2^+$, H$_3^+$, and their deuterated isotopologues
\citep[see][]{Flower2006, Taquet2013a}.
Deuterium fractionation reactions of the H$_3^+$-H$_2$ system have
been introduced in the network with the reaction rate coefficients
recently computed  by \citet{Hugo2009}. Deuteration reactions of the important
H$_2^+$-H$_2$, H$_3^+$-H, H$^+$-H$_2$, CH$_3^+$-H$_2$, and 
C$_2$H$_2^+$-H$_2$ systems and recombinations on electronegative
charged grains have also been included following \citet{Roberts2000,
  Roberts2003, Roberts2004, Walmsley2004} and \citet{Pagani2009}.  
Finally, we considered the chemical network involved in the formation of
deuterated water at warm temperatures and including neutral-neutral 
and radical-neutral reactions, following \citet{Thi2010}. 
{We therefore added about 3000 reactions to the deuterated KIDA
  network, resulting in a gas phase chemical network of 15260 reactions.}

\subsubsection{Surface chemical network}

The formation of the main ice species (H$_2$O, CO$_2$, H$_2$CO,
CH$_3$OH, NH$_3$, CH$_4$) along with their deuterated isotopologues was
followed with the chemical network developed by
\citet{Taquet2013a} from laboratory experiments showing the
efficient formation of interstellar ice analogs on cold surfaces. 
The transmission probability of chemical reactions with activation
energy barriers included in the water and carbon dioxide networks as
well as the key reaction CO+H have been computed with the Eckart 
model. Transmission probabilities of addition, substitution, and
abstraction reactions included in the methanol network have been
deduced from the rates experimentally derived by \citet{Hidaka2009}. 

We also added the surface network introduced by
\citet{Garrod2008} to form complex organics. Primary
radical-primary radical, primary radical-secondary radical reactions,
and reactions between aldehydes and H atoms have been incorporated to
our network and extended to include their deuterated counterparts by
assuming statistical branching ratios. The transmission probabilities
through quantum tunneling of the aldehyde-H reactions have been
computed by assuming rectangular energy barriers with a width $a = 1$
\AA. 
In total, the gas-grain chemical network includes about 2700 species and
24500 chemical processes.

For each species $i$, the effective binding energy $E_b(i)$ relative
to the surface was derived from the fractional coverage of $i$, $P(i)$
and H$_2$, $P$(H$_2$) on the surface layer and from the fraction of
bare surface, $P_{\textrm{bare}}$ lying under the surface layer,
following this formula: 
\begin{eqnarray}
E_b (i)  = & (& 1-P(\textrm{H}_2)-P(i)-P_{\textrm{bare}}) \times
E_{b,\textrm{wat}}(i)   \\ 
 & + & P(\textrm{H}_2) \times E_{b,\textrm{H}2}(i) + P_{\textrm{bare}} \times
E_{b,\textrm{bare}}(i) \\
 & + & P(i) \times E_{b,\textrm{pure}}(i) \nonumber 
\label{EbH} 
\end{eqnarray}
where $E_{b,\textrm{wat}}(i)$, $E_{b,\textrm{H}2}(i)$,
$E_{b,\textrm{bare}}(i)$, and $E_{b,\textrm{pure}}(i)$ are the binding
energies of species $i$ relative to amorphous solid water (ASW) ice,
H$_2$ ice, a bare (amorphous carbon or silicate) substrate and a
pure ice. 
To deduce $E_{b,\textrm{H}2}$ for all species, we applied the scaling
factor
\begin{equation}
E_{b,\textrm{H}2}(i) = E_{b,\textrm{H}2}(\textrm{H}) \times 
\frac{E_{b,\textrm{wat}}(i)}{E_{b,\textrm{wat}}(\textrm{H})}
\end{equation}
where $E_{b,\textrm{H}2}(\textrm{H})$ is 45 K \citep{Vidali1991}.
Table \ref{listEb} lists the sets of binding energies of selected
species for different substrates. In spite of their higher mass,
deuterated isotopologues seems to show similar binding energies than their main
isotopologue \citep[see][for a discussion of the binding energy
values]{Taquet2013a}. We therefore assumed the same binding energies
for all deuterated species and their main isotopologue.

\subsubsection{Input parameters} \label{input_params}

The {initial elemental abundances in the gas phase} considered in this
work are listed in Table \ref{elem_abu}. The elemental abundances of
species other than He and HD correspond to the set EA1 from
\cite{Wakelam2008}. The initial abundances of He and HD are taken from
the set EA2 from \citet{Wakelam2008} and from \citet{Linsky2003},
respectively.

\begin{table}[htp]
\centering
\caption{{Initial elemental abundances in the gas phase} with respect to hydrogen nuclei.}
\begin{tabular}{l c }
\hline
\hline
Species & Abundance \\
\hline
H$_2$ & 0.5 \\
HD & $1.6 \times 10^{-5}$ \\
He & $9.0 \times 10^{-2}$ \\
C & $7.3 \times 10^{-5}$ \\
N & $2.1 \times 10^{-5}$ \\
O & $1.8 \times 10^{-4}$ \\
Si & $8.0 \times 10^{-9}$ \\
S & $8.0 \times 10^{-8}$ \\
Fe & $3.0 \times 10^{-9}$ \\
Na & $2.0 \times 10^{-9}$ \\
Mg & $7.0 \times 10^{-9}$ \\
Cl & $1.0 \times 10^{-9}$ \\
\hline
\end{tabular}
\label{elem_abu}
\end{table}

The value of the ortho/para ratio of H$_2$ in molecular clouds is still highly
uncertain. The initial value of H$_2$ opr formed on grain surfaces is
most likely 3, as recently confirmed by the experiment by
\citet{Watanabe2010} conducted on amorphous solid
water. However, proton-exchange reactions in the gas phase would then
convert ortho-H$_2$ to para-H$_2$, decreasing the opr(H$_2$) towards the
Boltzmann value \citep[$\sim 3 \times 10^{-7}$ at 10 K,
see][]{Flower2006}.  
\citet{Pagani2011} interpreted the non-detection of DCO$^+$ towards
dark cloud envelopes with an opr(H$_2$) higher than 0.1. 
However, indirect estimates of H$_2$ based on the
comparison between observations of dark cores or cold protostellar
envelopes with chemical models suggest lower values of about
$10^{-3}-10^{-2}$ \citep{Pagani2009, Dislaire2012, Legal2013}.
In this work, the standard value for the initial opr(H$_2$) is set
to $10^{-2}$. This value corresponds to an average between
the values derived in the early stages (dark cloud envelopes) and the
late stages (protostellar envelopes) of molecular evolution.
Nevertheless, given the relative uncertainty of this 
value and its importance in the molecular deuteration process, we
also studied the effect of other initial opr(H$_2$) values ($10^{-4}$
and 3) on the ice deuteration.

We aimed at investigating the influence of some physical parameters on
the formation and the deuteration of interstellar ices. Therefore, we
assumed several values for the timescale of static formation, initial
density, visual extinction and temperature at the edge of the core,
listed in Table \ref{freeparams}. We also considered an other core,
model B, aiming at modelling the well-known core L1544, showing a higher
central density $n_{\textrm{H,0,obs}} = 4 \times 10^6$ cm$^{-3}$, a smaller
characteristic radius $R_{\textrm{f}} = 2 \times 10^3$ AU, and a
shallower slope $\eta = 2.5$ \citep{Crapsi2007} than L1498. 

As discussed in \citet{Taquet2012a} and \citet{Taquet2013a}, the
binding energy of light particles $E_b$ and the diffusion to binding
energy ratio $E_d/E_b$ show distributions of values depending on the
ice composition and morphology. We, therefore, explored several values
for these two parameters in a range given by the literature.

We kept fixed the following other parameters: \\
- distance between two binding sites $d_s = 3.1 $ \AA; \\ 
- cosmic-ray ionization rate $\zeta = 3 \times 10^{-17}$ s$^{-1}$; \\ 
- interstellar radiation field (ISRF) $F_{\textrm{ISRF}} = 1 \times 10^8$
photons cm$^{-2}$ s$^{-1}$; \\ 
- cosmic ray induced radiation field (CRH2RF) $F_{\textrm{CRH2RF}} = 5 \times
10^4$ photons cm$^{-2}$ s$^{-1}$; \\ 
- scaling factor in multiples of the local interstellar field $G_0 = 10$.


\begin{table}[htp]
\centering
\caption{List of free input parameters and their value range explored in
  this work.}
\begin{tabular}{l c}
\hline
\hline
Input parameters & Values \\
\hline
Type of core & {\bf A} - B (see text) \\
$n_{\textrm{H,ini}}$ (cm$^{-3}$) & $\mathbf{3\times 10^3}$ -$10^4$ \\
$A_{\textrm{V,ext}}$ (mag) & 1 - {\bf 2} \\
$T_{\textrm{d,ext}}$ (K) & 14 - {\bf 17} - 20 \\
$t_\textrm{BC}$ (yr) & $\mathbf{1} - 2 \times t_{\textrm{BC}}$ \\
o/p(H$_2$)$_{\textrm{ini}}$ & $1 \times 10^{-4}$ - $\mathbf{1 \times 10^{-2}}$ -
3 \\
$E_d/E_b$ & 0.5 - {\bf 0.65} - 0.8 \\
$E_b$(H) (K) & 400 - {\bf 500} - 600 \\
\hline
\end{tabular}
\tablecomments{Bold values mark the values adopted in the reference
  model.}
\label{freeparams}
\end{table}


\section{Results}

\subsection{Physical model validation}

\begin{figure*}[htp]
\centering
\includegraphics[width=80mm]{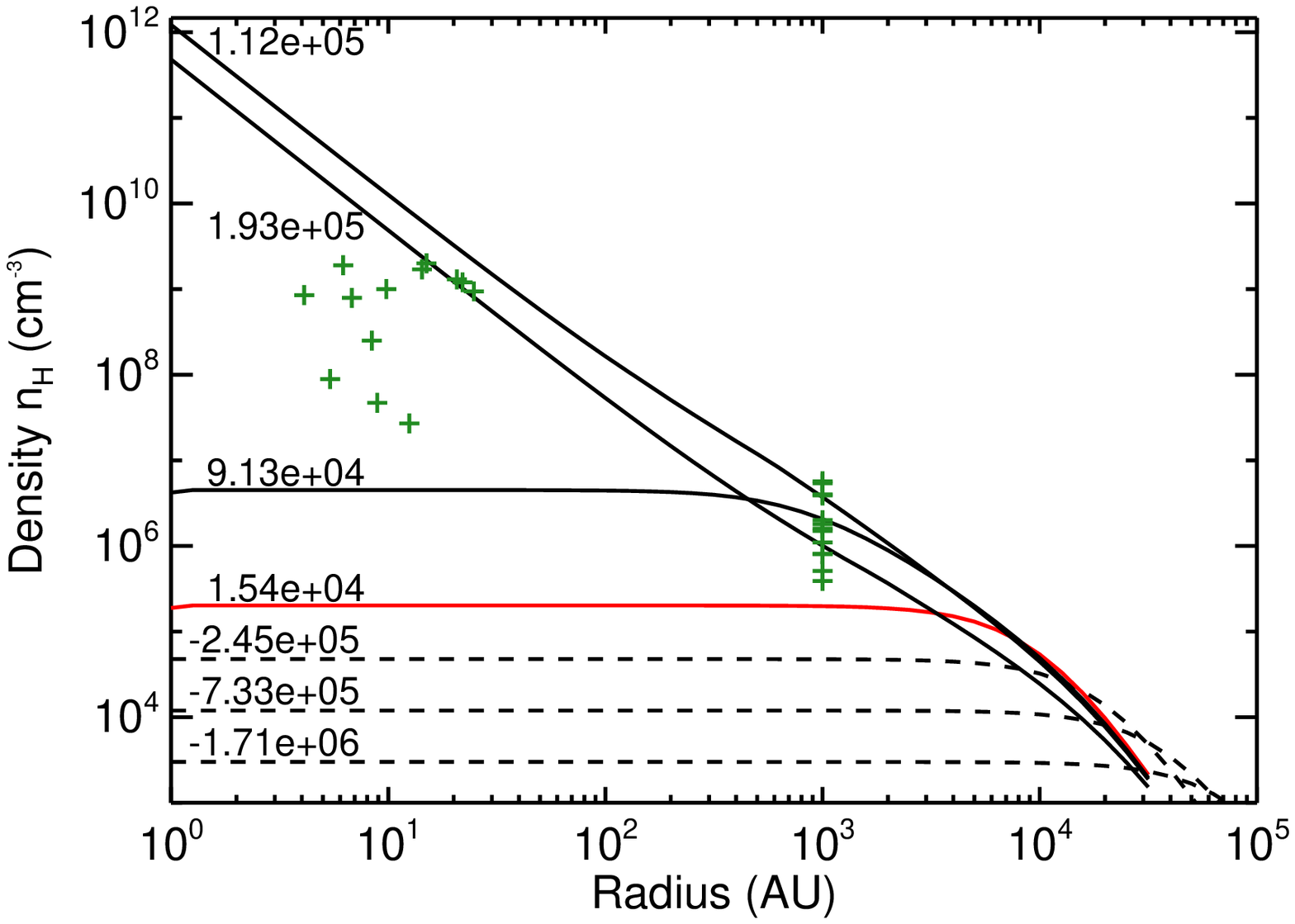}
\includegraphics[width=80mm]{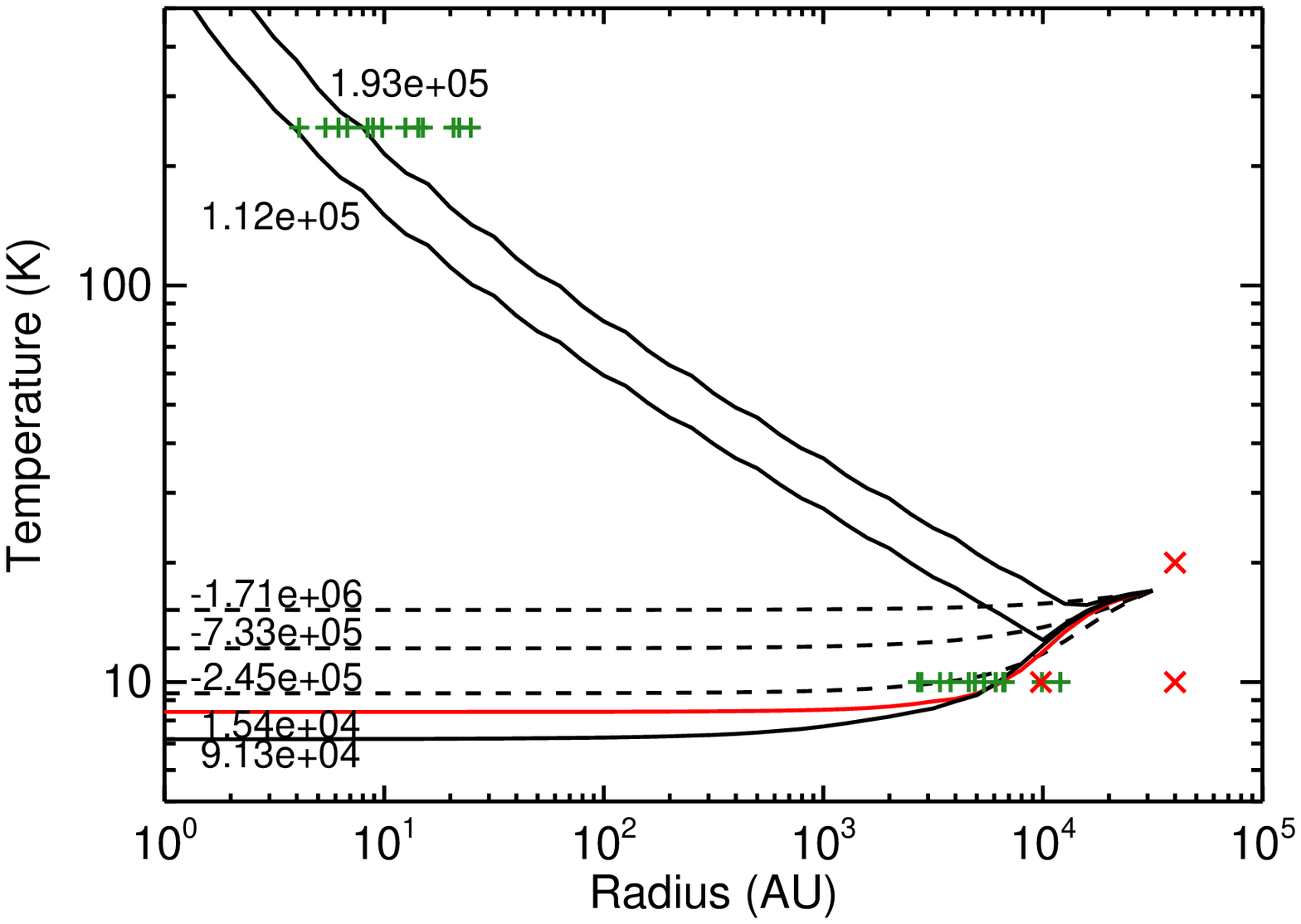}
\includegraphics[width=80mm]{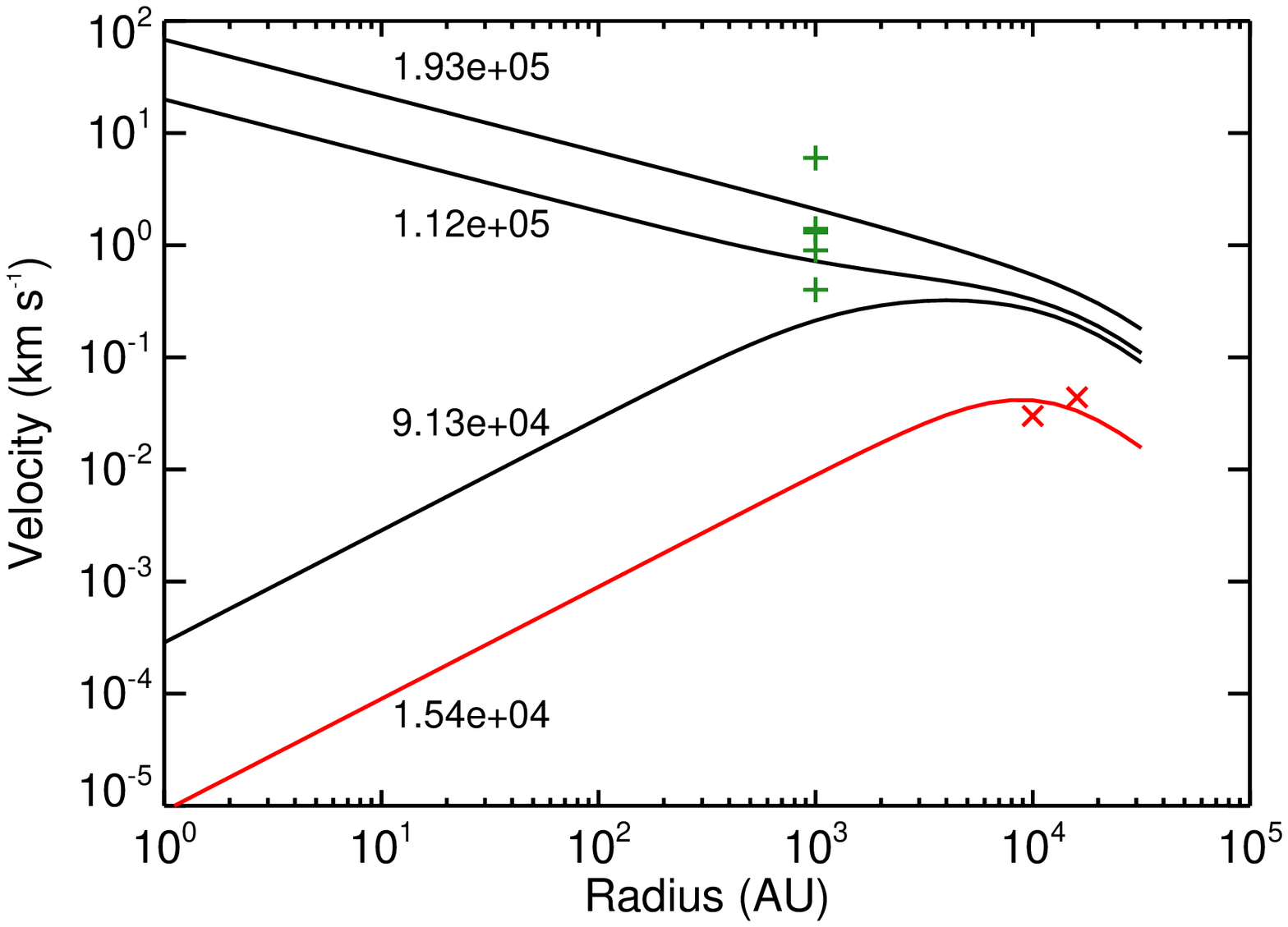}
\includegraphics[width=80mm]{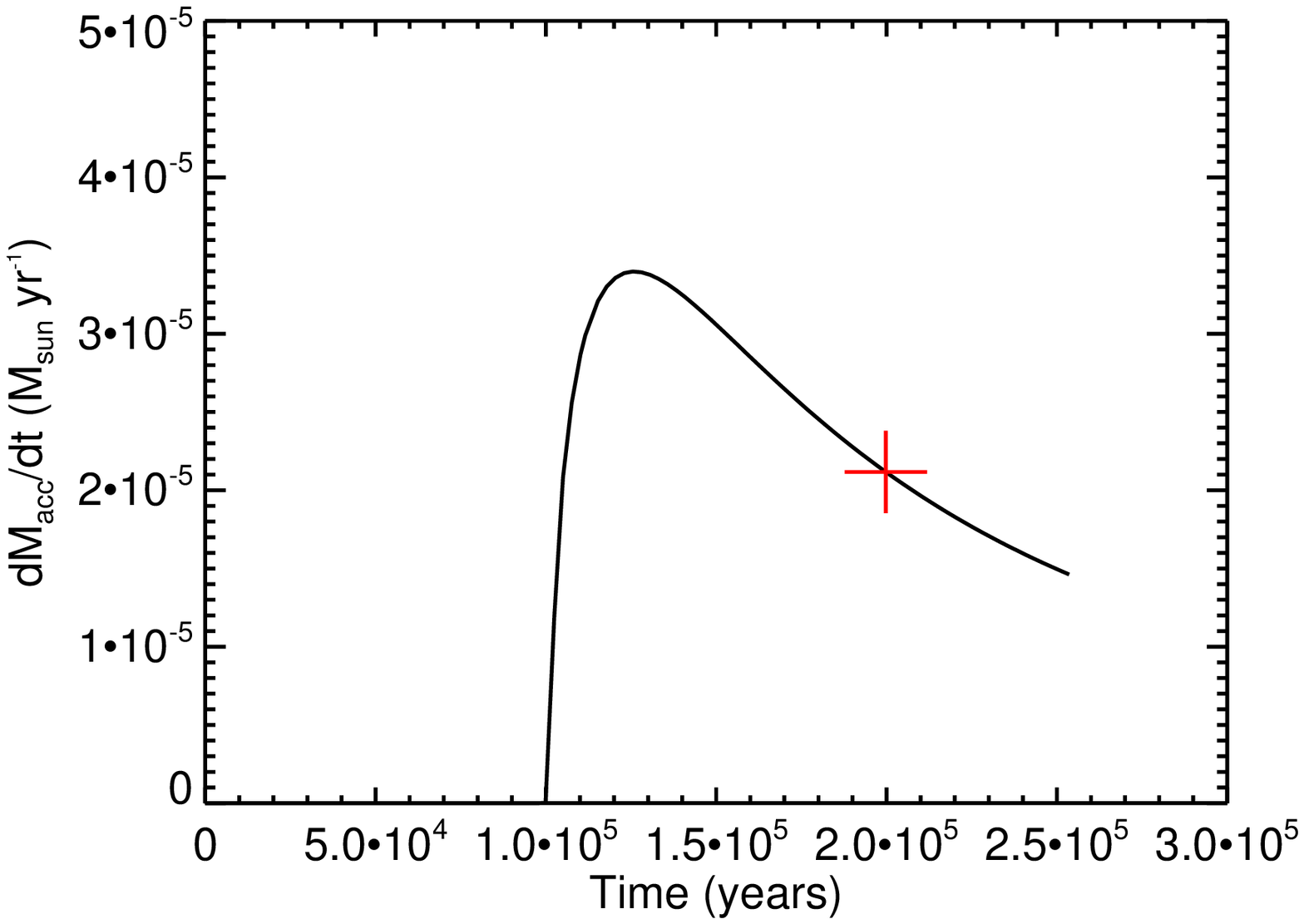}
\caption{Density (top left), temperature (top right), and infall
  velocity (bottom left) profiles of model A-F plotted for different
  times and overplotted with observational data of prestellar (red crosses) and
  protostellar (green plus) cores (see text for more details).
 Bottom right: Evolution of the accretion rate as a function of
 time. The red cross represents the end of the Class 0 stage defined
 as the time when the mass of the protostar reaches half of the total
 (protostar+envelope) mass.} 
\label{phys_results}
\end{figure*}

Figure \ref{phys_results} presents the physical (density,
temperature, velocity) profiles and the accretion rate of the modelled
core A-F as a function of time, from the beginning of the translucent
phase to the end of the Class 0 stage.  
$t = 0$ is defined as the beginning of the core collapse, negative
times stand for the static formation of the dense core while positive
times correspond the dynamical collapse of the core. 

For $0 < t < t_{\textrm{FF}}$, the core shows slow internal motions that
decrease proportionally with the radius towards the center.   
The core keeps its Plummer-like structure, the density profiles are flat
in the center at $r < R_f$ and steepen towards the edge. However,
with time, the central density slowly increases while $R_f$ decreases. 
When one free-fall time is reached, all the shells located inside
$R_f$ collapse almost simultaneously because of their similar
free-fall timescale due to their nearly uniform density. It,
therefore, induces a quick increase of the accretion rate, that allows
the formation of the central protostar.  
The density profile quickly steepens at the center, the density at 10
AU increasing from less than $5 \times 10^6$ to $10^{10}$ cm$^{-3}$ in
$2 \times 10^4$ yr, and follows an asymptote with a power-law index of
$\sim 1.5$.  
Since the temperature profile increases with the protostar luminosity
and the protostar mass, the temperature in the protostellar envelope
also quickly increases from less to 10 K to 200 K at 10 AU.
After the formation of the central protostar, the accretion rate
slowly decreases because of the slow and gradual collapse of each
shell located outside $R_f$. 

In spite of its relative simplicity, the physical model used in this
work is in fair agreement with the observations of molecular clouds,
dense cores, and Class 0 protostars regarding their timescales and
their physical profiles, as shown in Table \ref{core_param} and Figure
\ref{phys_results}. 
Table \ref{core_param} compares the lifetimes derived from the
observations with the timescales obtained with the physical model. 
It can be seen that all the timescales obtained with the model lie in
the range of values deduced from the observations, derived as explained
below. 

By analysing the IRAS survey carried out by \citet{Beichman1986},
\citet{WardThompson1994} estimated a total lifetime of $\sim 1$ Myr
for prestellar cores detected in the submillimeter continuum. The
more recent c2d \textit{Spitzer} legacy project has yielded shorter
lifetimes.  \citet{Enoch2008} and \citet{Evans2009} estimated the
lifetime of prestellar cores to be $\sim 0.45$ Myr with variation from
cloud to cloud between 0.23 Myr in the Serpens cloud and 0.78 Myr in
Perseus. 
A significant fraction of cores displays internal motions of
contractions. The analysis of the line profiles of $\sim 200$ sources by
\citet{Lee1999} has shown that 3 \%  of them display contraction
signatures. However, more recent studies by \citet{Gregersen2000} and
\citet{Lee2004b} show that about 10-20 \% of them are contracting. 
Assuming a total prestellar core lifetime between 0.23 and 0.78 Myr
gives a lifetime of static cores $t_{\textrm{SC}}$ and contracting 
cores $t_{\textrm{CC}}$ of $2 \times 10^5 - 7 \times 10^5$ and $2
\times 10^4 - 1.6 \times 10^5$ yr, respectively. 
By comparing the number of Young Stellar Objects in different stages
detected in various clouds, \citet{Evans2009} and \citet{Maury2011}
estimated the lifetime of Class 0 protostars $t_{\textrm{C0}}$ of $1 -
1.6 \times 10^5$ yr and $0.4 - 1 \times 10^5$ yr, respectively.

Figure \ref{phys_results} compares the physical profiles obtained with
our model with observations of dark cores and low-mass protostars.
Red crosses represent the temperatures and velocities obtained towards
L1498 by \citet{Lee2001}, \citet{WardThompson2002}, and
\citet{Tafalla2004}  from observations of  CS, N$_2$H$^+$ and
C$^{18}$O transitions, from continuum observations obtained with {\it
  ISO} and  from the modelling of the emission of NH$_3$ line
transitions assuming a thermal coupling between dust and gas,
respectively.   
As can be seen in Fig. \ref{phys_results}, the temperature and
velocity profiles obtained with our model are in fair agreement with
the observations at the edge of the core.

Green plus symbols in Figure \ref{phys_results} stand for the density,
temperature, and velocity values derived by \citet{Kristensen2012} and
\citet{Mottram2013} for a sample of low-mass protostars observed by
the {\it Herschel Space  Observatory}. For each model, only the
observed protostars whose envelope masses are lower than the mass of
the modelled core are shown.  
The density profiles obtained with our model are in good agreement
with the observational values at 1000 AU while they are slightly
higher than the values observationally derived at the inner regions
($< 50$ AU) of the protostellar envelopes. 
The discrepancy probably results from the break of the 1D symmetry in
the inner 100 AU and the eventual formation of protoplanetary disks
during the Class 0 phase \citep[see][for instance]{Pineda2012}.   
The temperatures of the warm inner region reached at the end of the
Class 0 phase lie in the range of the observed values while the
predicted temperatures at the external part of the envelope seem to be
slightly higher than the observations, probably induced by the
coupling between the dust and gas temperatures assumed in our model.
The velocity profiles in the external protostellar envelopes obtained
with our model assuming a free-fall collapse lie in the range of the
values derived observationally. However, the 1D free-fall collapse
assumed in our model is not necessarily correct in the inner few
hundreds of AU and would tend to overestimate the infall velocity
occurring in the hot corino.

\subsection{Chemical differentiation of ices in dark clouds}

\subsubsection{Reference model}

\begin{figure}[htp]
\centering 
\includegraphics[width=85mm]{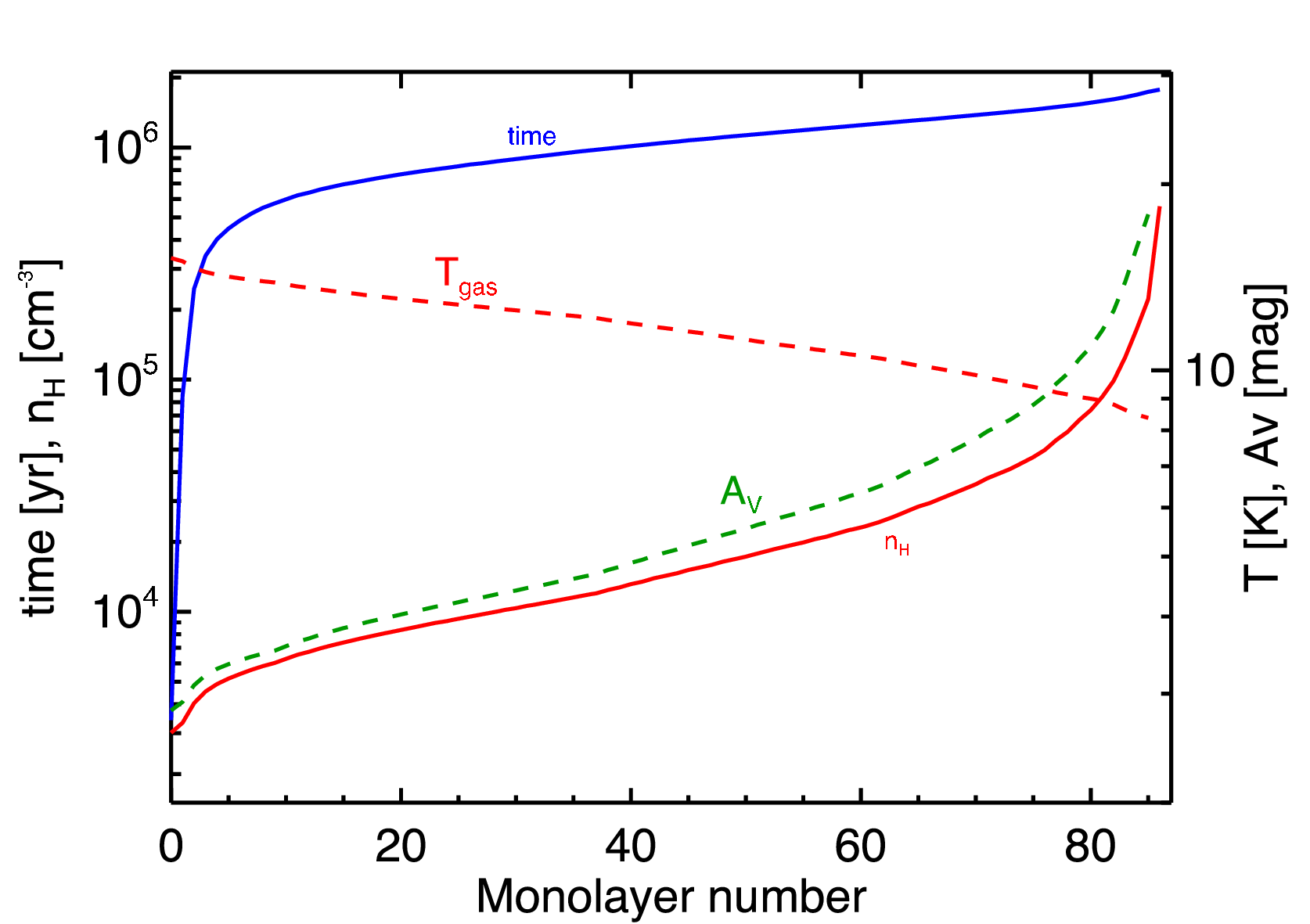}
\caption{Evolution of the physical properties as function of the
  number of layers formed in the ices obtained for a shell at the
  center of model A-F.}
\label{physprop_layers}
\end{figure}

\begin{figure*}[htp]
\centering 
\includegraphics[width=180mm]{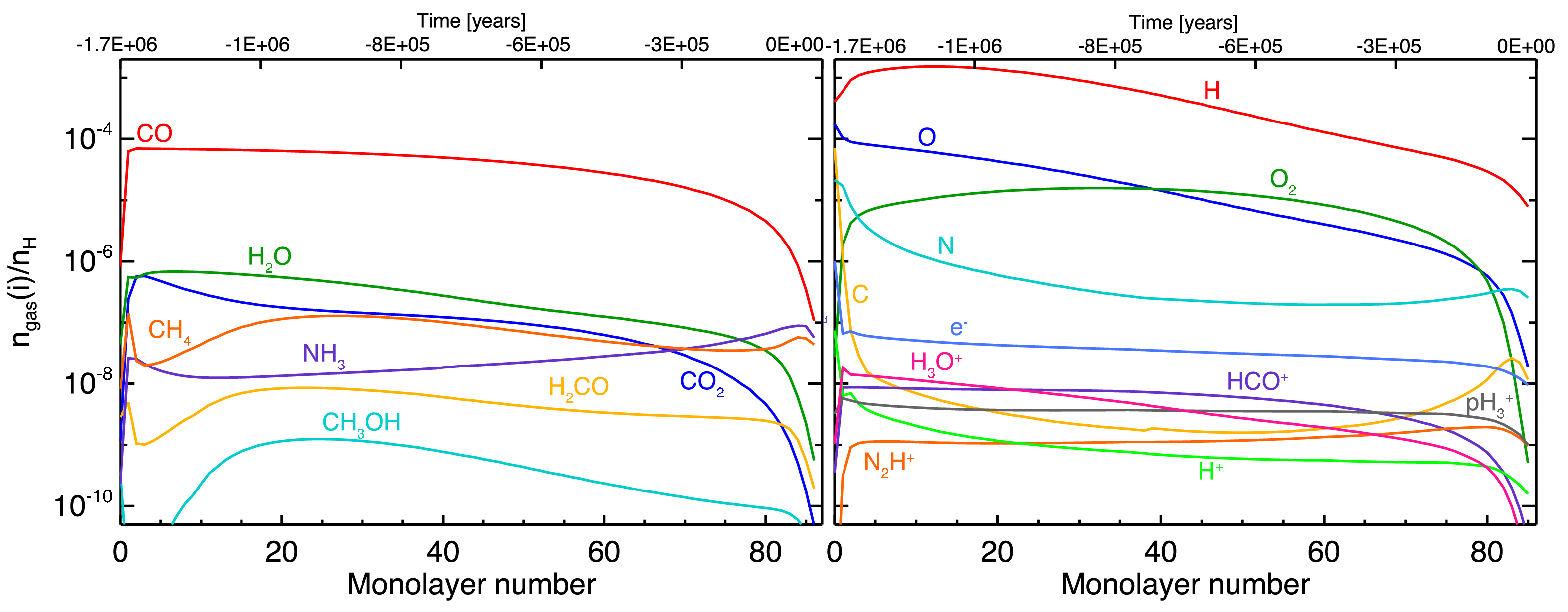}
\caption{Evolution of the gas phase abundances of important species as
  function of the number of layers formed in the ices obtained for a
  shell at the center of model A-F.}  
\label{X_layers}
\end{figure*}

\begin{figure}[htp]
\centering 
\includegraphics[width=90mm]{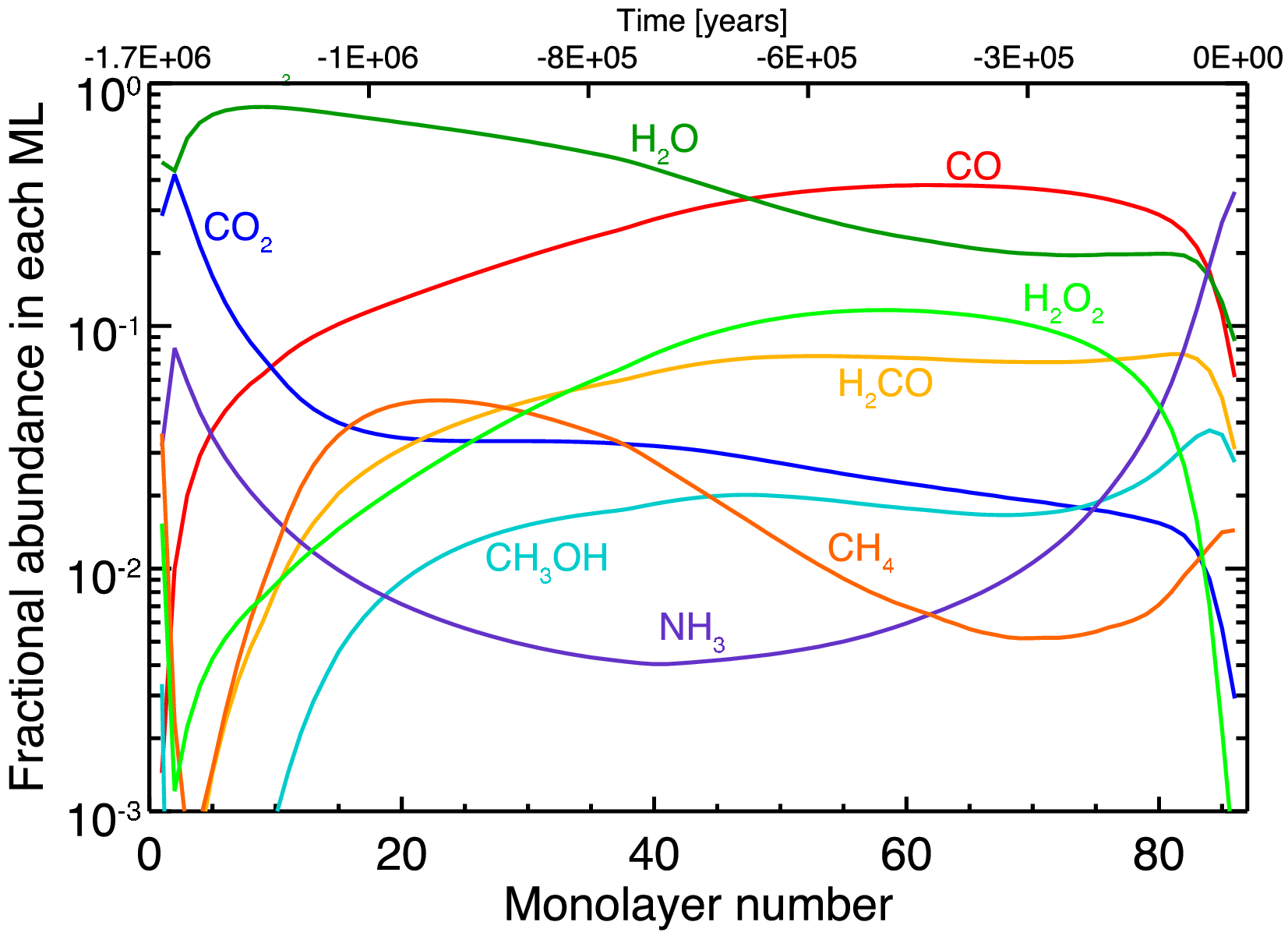}
\caption{Fractional composition in each ice monolayer as function of
  the number of layers formed in the ices obtained for a shell at the
  center of model A-F.}  
\label{poplay_layers}
\end{figure}

\begin{figure*}[htp]
\centering 
\includegraphics[width=180mm]{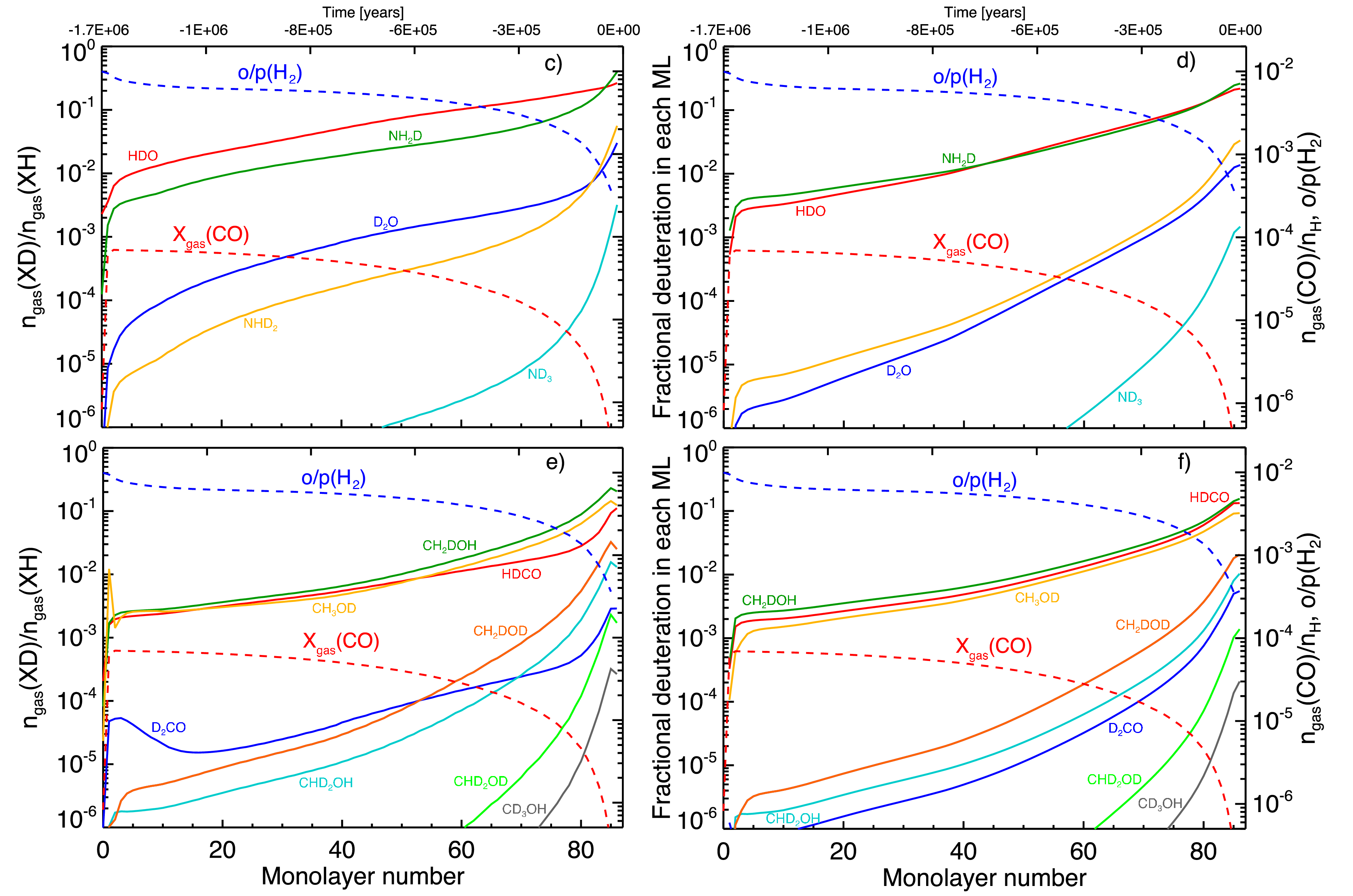}
\caption{Evolution of the gas phase deuteration (left) and the icy
  deuteration in each layer (right) for water and ammonia (top) and
  for formaldehyde and methanol (bottom) as function of the number of
  layers formed in the ices obtained for a shell at the center of
  model A-F. The evolution of the gaseous CO abundance and the ortho/para
  ratio of H$_2$ are overplotted in dashed lines and their values are shown on the right axis.} 
\label{D_layers}
\end{figure*}

The physical model used in this work allows us to follow the
multilayer formation of interstellar ices from the sparse translucent
phase of molecular clouds to the dense and cold phase seen at the
center of dark cores. 
Figures \ref{physprop_layers}, \ref{X_layers}, \ref{poplay_layers},
and \ref{D_layers} present the evolution of the physical and chemical
properties of a shell located at the center of  the model A-F as a
function of the ice deposition profile. The time corresponding to any
layer represents the time at which the layer has been formed.  

Fig. \ref{physprop_layers} shows the evolution of the physical
properties ($n_{\textrm{H}}$, $T$, $A_{{V}}$) and time with the formation of
  ices. Most of the ice formation occurs in molecular clouds since
  about 80 \% of the ice monolayers are formed when $n_{\textrm{H}} <
  5 \times 10^4$ cm$^{-3}$ and $A_{V,\textrm{mod}} < 10$ mag. 
Consequently, only 20 \% of the ices has been formed in cold
($T < 10$ K) and dense ($n_{\textrm{H}} > 5 \times 10^4$ cm$^{-3}$)
conditions.

The evolution of the physical conditions induces an evolution of the 
gas phase chemistry shown in Fig. \ref{X_layers} for a selection of
important gaseous species and for the main ice components. 
In particular, the abundances of atomic C, O, and N do not display the
same evolution. Atomic C is quickly transferred to CO via reactions in
the gas phase while the abundance of atomic O slowly decreases because of
its freezes-out onto the grain surfaces forming icy species, or
through reactions in the gas phase forming gaseous species such as
O$_2$. The modelled abundance of O$_2$ is several orders of magnitude
higher than the observations of molecular clouds carried out with
space telescopes \citep[see][]{Larsson2007, Liseau2012}.
Atomic nitrogen shows a decrease of its abundance due to
its freeze-out on the grains and through reactions in the gas phase.
Since CO is the main destroyer of N$_2$H$^+$, its depletion in dense
core conditions tends to re-increase the abundance of N-bearing
species, such as N$_2$H$^+$, NH$_3$ or N, at the end of the
calculation. This effect has already been
observed by \citet{Bergin2001} and \citet{Tafalla2002} and previously
modelled by \citet{Bergin1997} and \citet{Aikawa2001}.
The gas phase abundances of the main icy components lie within a few
orders of magnitude since some species (CH$_3$OH) are only formed at
the surface of grains and are then evaporated through non-thermal
processes while others (H$_2$O, H$_2$CO, NH$_3$) can also be formed in
the gas phase and show higher abundances. 

The mantle of cold interstellar ices is believed to be mostly inert
\citep{Watanabe2004}. The chemical composition of each layer would,
therefore, reflect the physical and chemical properties at the moment
of its formation. 
Figure \ref{poplay_layers} shows the chemical composition of each
layer within interstellar ices formed with our reference model.
It can be seen that the ices are very heterogeneous and their
chemical structure depends on the evolution of both the physical
properties and the gas phase abundances \citep[see also][]{Garrod2011,
  Taquet2013a, Vasyunin2013}. 
Water and carbon dioxide are mostly formed in the inner parts of ices when the
density is limited with high atomic abundances, and
when the temperature is warm, allowing the diffusion of heavier
species, such as CO, O, or OH, to recombine together to form CO$_2$.
As the density increases and the temperature decreases with time, CO
freezes-out onto the grains and is gradually destroyed to form H$_2$CO and
CH$_3$OH instead of CO$_2$.
Interestingly, NH$_3$ shows high abundances in the inner and outer
parts of the ices. The high abundance in the inner part is due to the
high abundances of H and N that favour the hydrogenation of atomic N
on grains while the high abundance at the surface of the ice is due to
the increase of N-bearing species in the gas phase at high CO depletions.

As mentioned in the introduction, the deuteration of gaseous and icy
species depends on several physical (density, temperature) and
chemical (opr(H$_2$), $X_{gas}$(CO)) parameters.
Hydrogenated icy species are mostly formed through
hydrogenation reactions triggered by the irradiation of H atoms at the
surface of interstellar grains. The atomic [D]/[H] abundance ratio in
the gas phase, therefore, governs the deuteration on the ices. 

The effect of these two chemical parameters on the deuteration can be
seen in Fig. \ref{D_layers} showing the evolution of the deuteration in the
gas phase and in each ice monolayer, overplotted with the evolution of
opr(H$_2$) and $X_{gas}$(CO), whose values are shown on the right
axis, as function of the ice deposition profile. 
The timescale needed to decrease opr(H$_2$) and $X_{gas}$(CO) are
similar to the timescale of the total ice formation. 
The two parameters slowly decrease during most of the ice formation,
keeping high values ($> 10^{-3}$ and $> 10^{-5}$, respectively) up to
70-80 MLs,  inducing a slight increase of the deuteration both in the
gas phase and on ices.  
The decrease of these two crucial parameters is enhanced at the end of
the ice formation for dense core conditions, and generates a sharp increase
of the deuteration of doubly and even more triply deuterated species
near the surface. 
It can be seen that the deuteration of water, ammonia, formaldehyde, and
methanol in the gas phase reflects the deuteration within each
monolayer which, in turn, roughly follows the evolution of 
the atomic D/H abundance ratio in the gas phase, reaching values
higher than 10 \% at the end of ice formation.
Doubly deuterated species show fractionations higher than 1 \% while
triply deuterated forms of methanol show D/H ratios higher than
$10^{-4}$ at the surface. 

The deuteration of gaseous water seems to be slightly enhanced
compared to the evolution of the atomic D/H ratio and the deuteration
of formaldehyde and methanol. It is due to its efficient deuteration
in the gas phase through ion-neutral reactions, as also predicted by
the gas phase calculations of \citet{Roberts2004}.  
CH$_2$DOH and CH$_3$OD show similar deuteration profiles but their final
abundances in the outermost layers differ by $\sim 60$ \%. The
difference is due to the abstraction reactions occurring only on the methyl
group of methanol introduced in our chemical network following the
experiments by \citet{Hidaka2009}.

\subsubsection{Influence of input parameters} \label{influence_params}

We have investigated the influence of several physical and chemical
parameters on the formation and the deuteration of interstellar
ices. For this purpose, we varied the values of the parameters listed
in Table \ref{freeparams} to our reference model. Table
\ref{params_ice} and Table \ref{params_deut} of the Appendix
present the ice composition and the deuteration of water,
formaldehyde, and methanol obtained when the density reaches the
observed value for the considered core  (L1498: $n_{\textrm{H,0,obs}} = 2 \times
10^5$ cm$^{-3}$ for model A and  L1544: $n_{\textrm{H,0,obs}} = 4 \times 10^6$
cm$^{-3}$  for model B).  

The variation of the five physical properties (type of core, initial density,
initial external visual extinction, initial external temperature, and
timescale for the static formation) slightly influence the chemical
composition of ices, in particular the conversion from CO to CO$_2$ and to
H$_2$CO and CH$_3$OH. 
CO$_2$ is mainly formed at the beginning of the molecular cloud phase,
through reactions involving heavy particles, at low-density and warm
conditions ($T>12$ K) while H$_2$CO and CH$_3$OH are rather formed in the
external part of ices for dense and colder conditions. 
The increase of the initial density decreases the abundance of H atoms
in the gas phase and, therefore, the efficiency of hydrogenation
reactions forming formaldehyde and methanol in competition with the
H$_2$O and CH$_4$ formation.
The increase of the initial visual extinction decreases the
photodesorption of reactive particles and, therefore, increases the
efficiency of surface reactions. 
The increase of the temperature increases the mobility of heavy
species allowing the formation of CO$_2$, but also the mobility of H
atoms in dense core conditions when the H/CO abundance ratio is high,
enhancing the formation of formaldehyde and methanol. 

The surface parameters, namely the binding energies of light particles
and the diffusion to desorption energy ratio, strongly influence
the chemical composition of ices because the rates of surface
reactions depend exponentially on them. 
Therefore, the increase of $E_b$ and $E_d/E_b$ strongly decrease the
abundances of surface molecules formed through reactions with
activation energies (CO$_2$, H$_2$CO, and CH$_3$OH). These two
parameters are, therefore, the two most important parameters for the
chemical composition of ices. 
{By following the multilayer ice formation during a free-fall
  collapse with their Monte-Carlo model, \citet{Vasyunin2013} observed
  the same influence of the diffusion-to-binding energy ratio on the
  chemical composition of ices.}
%
The comparison of our modelling results with typical interstellar ice
abundances deduced from infrared observations of ices show that,
although the chemical composition of most species strongly depend on
$E_d/E_b$ and $E_b$,  the reference models give predictions that are in fair
agreement with observations. 
Overall, CO$_2$ tends to be slightly underproduced while H$_2$CO, CO, and
NH$_3$ show overpredictions of their abundances.

\begin{figure*}[htp]
\centering 
\includegraphics[width=180mm]{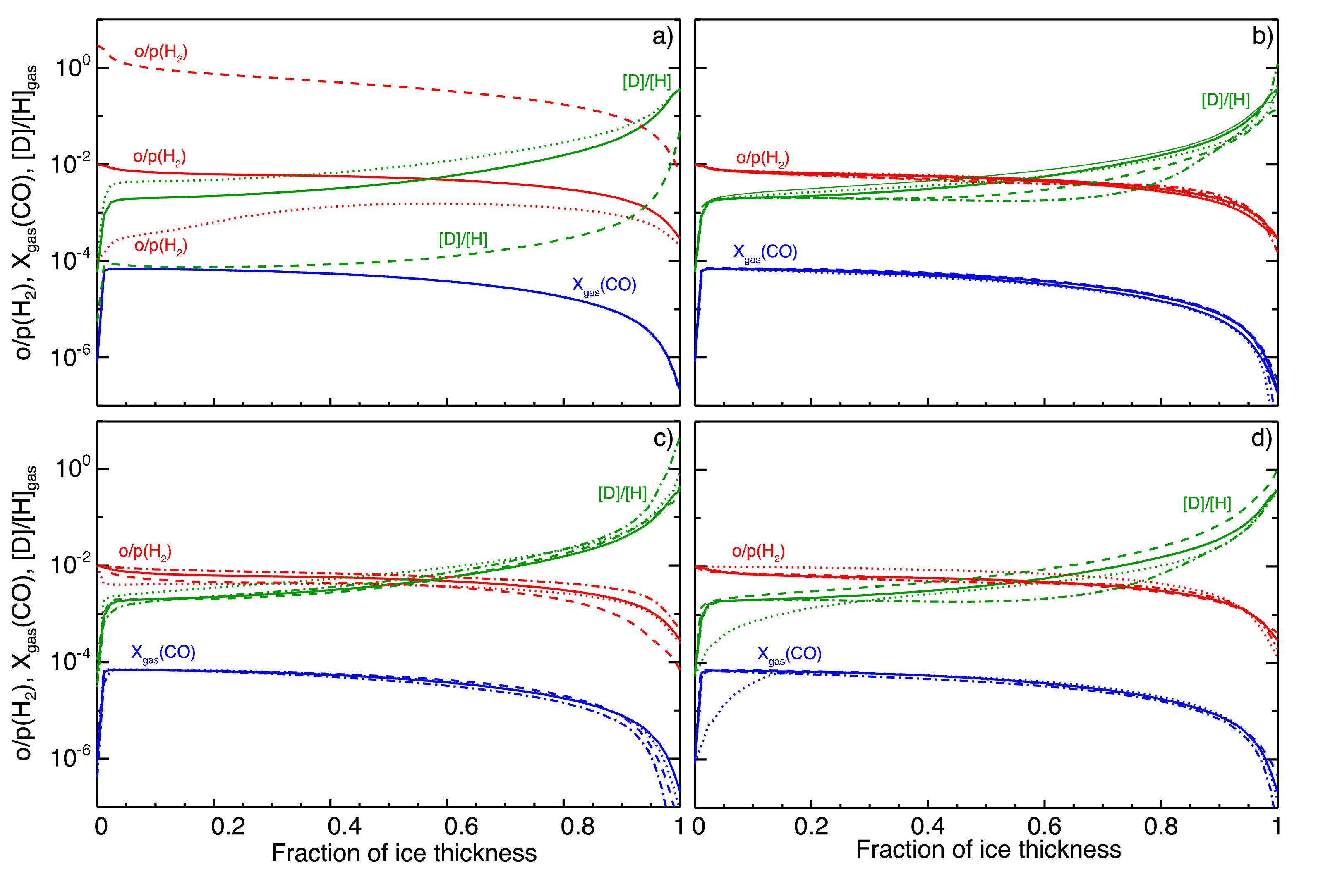}
\caption{Evolution of H$_2$ opr, gas phase abundance of CO, and
  gaseous atomic [D]/[H] abundance ratio as a function of the
  normalized thickness of interstellar ices. 
a) Solid lines: reference model, dotted lines: opr(H$_2$) = $10^{-4}$,
dashed lines: opr(H$_2$) = 3.
b) Solid lines: reference model, dotted lines: $E_d/E_b = 0.5$, dashed
lines: $E_d/E_b = 0.8$, dotted-dashed lines: $E_b$(H) $= 400$ K, thin
solid lines: $E_b$(H) $= 600$ K.
c) Solid lines: reference model, dotted lines: $A_V = 1$ mag, dashed
lines: $t_{\textrm{BC}} = 2 \times t_{\textrm{BC,0}}$, dotted-dashed
lines: model B.
d) Solid lines: reference model, dotted lines: $n_{\textrm{H,ini}} =
10^4$ cm$^{-3}$, dashed lines: $T_{\textrm{d,ext}} = 14$ K ,
dotted-dashed lines: $T_{\textrm{d,ext}} = 20$ K.} 
\label{inputparams_layers}
\end{figure*}

Not surprisingly, the most important parameter for the deuterium
chemistry is the initial value of the opr of H$_2$. 
Since the timescale needed to decrease the opr(H$_2$) is similar to the
timescale of total ice formation, a high initial opr(H$_2$) value of 3
gives a final opr(H$_2$) of $\sim 5 \times 10^{-2}$ 
(see Fig. \ref{inputparams_layers} showing the evolution of the opr of
H$_2$, the CO abundance in the gas phase, and the atomic gaseous D/H
abundance ratio as function of the normalized ice deposition profile
for each model). The deuteration
reactions are, therefore, strongly limited by the high abundance of
o-H$_2$ throughout the formation of dense cores, decreasing the
deuterium fractionation of reactive ions, such as the
p-D$_2$H$^+$/o-H$_2$D$^+$ and N$_2$D$^+$/N$_2$H$^+$ abundance ratios,
by more than one order of magnitude. 
Since the overall deuteration of icy species is governed by the atomic
D/H abundance ratio in the gas phase which is, in turn, mostly
governed by the deuteration of H$_3^+$ \citep{Roberts2003,
  Taquet2012b}, the deuteration in ices is also reduced by
more than one order of magnitude. 

The variation of other chemical and physical parameters has only a
moderate impact on the atomic gaseous [D]/[H] abundance ratio and,
therefore, on the ice deuteration since they do not influence the
variation of opr(H$_2$) and $X_{gas}$(CO)
as it can be seen in Fig. \ref{inputparams_layers}.
The three parameters roughly follow the same trend for all the
parameters explored in this work, except when the opr(H$_2$) is
varied. 
The deuteration in ices is, therefore, almost independent of these
parameters. The variation of the overall ice deuteration seen in Table
\ref{params_deut} is mainly due to the variation of their moment of
formation within ices. For instance, the increase of the initial
temperature increases the abundance of methanol which is also located
in the inner and less deuterated part of ices.

\subsection{Spatial evolution in dark cores}

\begin{figure*}[htp]
\centering 
\includegraphics[width=180mm]{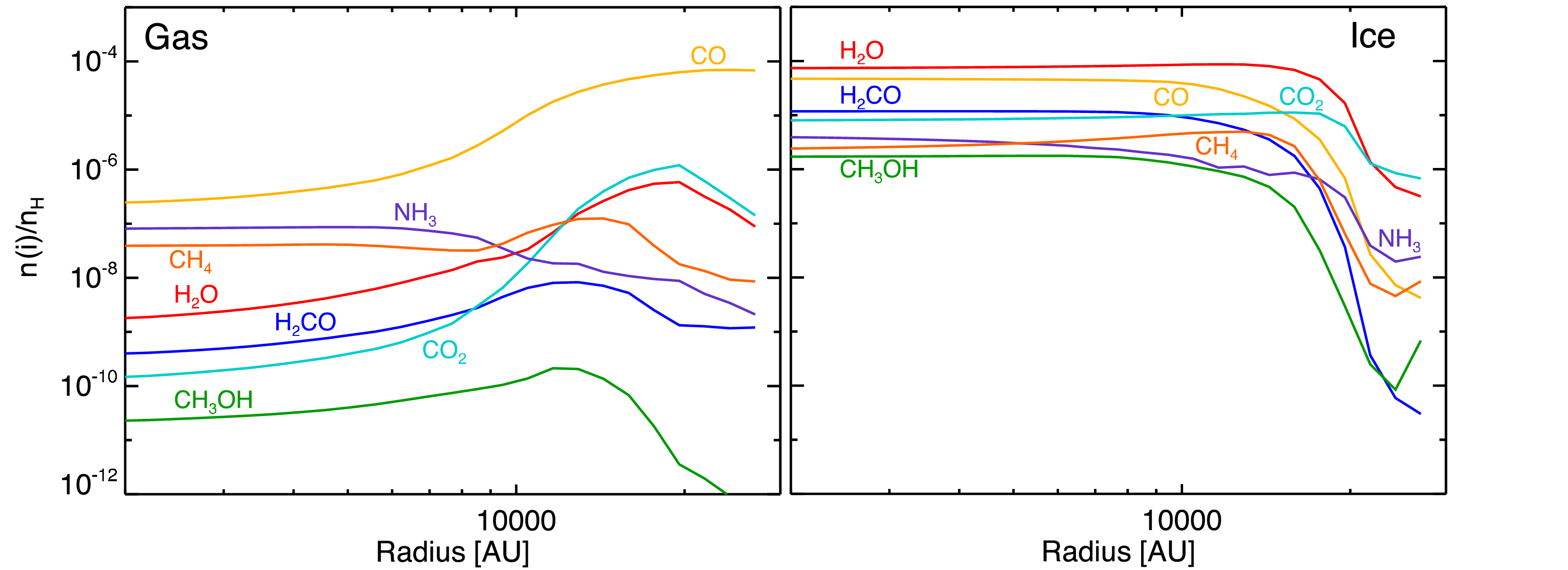}
\caption{Abundance profiles {in the gas phase (left) and in ices
    (right)} of the main ice components
  obtained for the L1498 model obtained  when the density profile reaches the observed profile.}
\label{L1498core_abu}
\end{figure*}

\begin{figure*}[htp]
\centering 
\includegraphics[width=180mm]{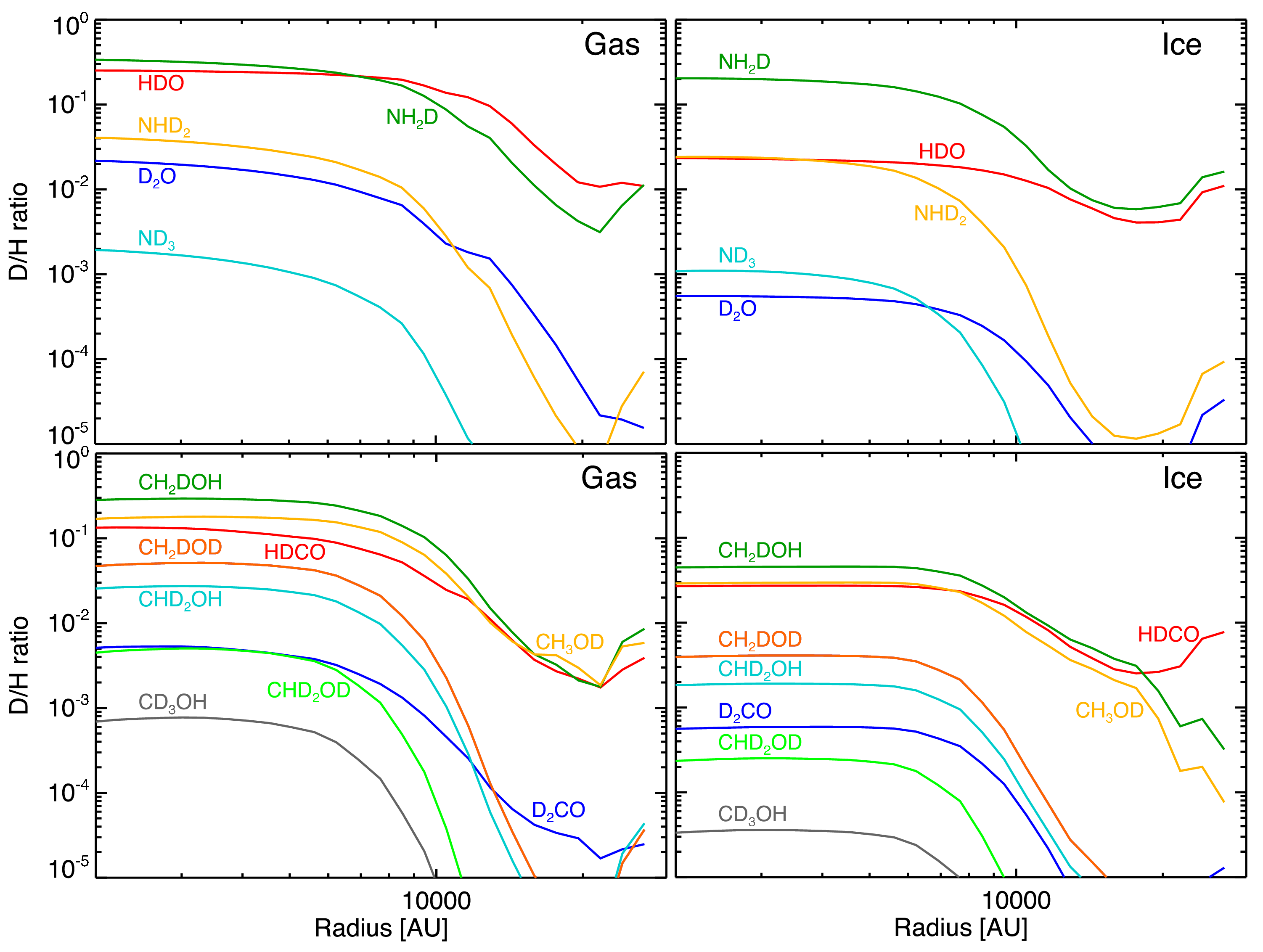}
\caption{Deuteration profiles {in the gas phase (left) and in ices
    (right)} of the main ice components
  obtained for model A-F obtained  when the density profile
  reaches the observed profile.} 
\label{L1498core_deut}
\end{figure*}

The abundances and deuterations profiles of the main ice
components in the gas phase and in ices, obtained when the density
profile reaches the observed profile of L1498, are shown
in Figures \ref{L1498core_abu} and \ref{L1498core_deut}. 
%
Every species shows a decrease of the abundance in ices towards
the edge of the core, mainly due to the high flux of UV photons which
photo-evaporates and photo-dissociates interstellar ices. 
However, the abundances of formaldehyde and methanol decrease more
strongly, by five orders of magnitude, and at lower radius ($r \sim 1.5
\times 10^4$ AU) than other species because of the warm temperature
reached at the edge of the core that favours the CO$_2$ formation from
CO instead of formaldehyde and methanol.  
%
%
Unlike interstellar ices, the abundances in the gas phase of most of icy
species tend to increase towards the edge of the core because they are
governed by the balance between their formation through the
evaporation from grains supported by an eventual formation in the gas
phase (for CO, H$_2$O, H$_2$CO, or NH$_3$) and their depletion through
the freeze-out on grains in the center of cores or their
photodissociation at the edge of the cores. 
{Thus, the gas-phase abundances of most of species tend to deplete towards the
center because of the increase of the density, enhancing the accretion
rate, and the decrease of the temperature that decreases the
evaporation rate.  }
Species that are more strongly bound to the ices and that
do not show any efficient formation in the gas phase, such as CO$_2$
and CH$_3$OH, display lower abundances at the center. In contrast, and
as already discussed previously, NH$_3$ abundance tends to increase
towards the center of the core because its formation in the gas phase
and on ices is enhanced by the CO depletion.

As shown in Fig. \ref{L1498core_deut}, the icy and gaseous
deuterations of all species increase towards the center of the core,
because of the increase of the density and the decrease of the
temperature which decrease $X_{gas}$(CO) and opr(H$_2$). 
The deuteration of all simply deuterated isotopologues is of a few
percents in ices except for the deuteration of ammonia which is about 20
\% because of its late formation. 
The deuteration of methanol decreases more strongly towards the edge
of the core than for other species because CH$_3$OH deuteration is not
efficient at warmer temperatures while gas phase reactions can still
account for the deuteration of H$_2$O, H$_2$CO, and NH$_3$. 
The deuteration of all icy species in the gas phase is higher than the
overall deuteration in ices by about one order of magnitude. This is
due to the deuteration differentiation in ices; the deuteration of the
gas phase roughly scales with the deuteration of the highly deuterated
surface of ices while the overall deuteration in ices also includes
the inner and poorly deuterated ice mantle.   

\subsection{Ice evaporation in Class 0 envelopes}

\begin{figure*}[htp]
\centering 
\includegraphics[width=180mm]{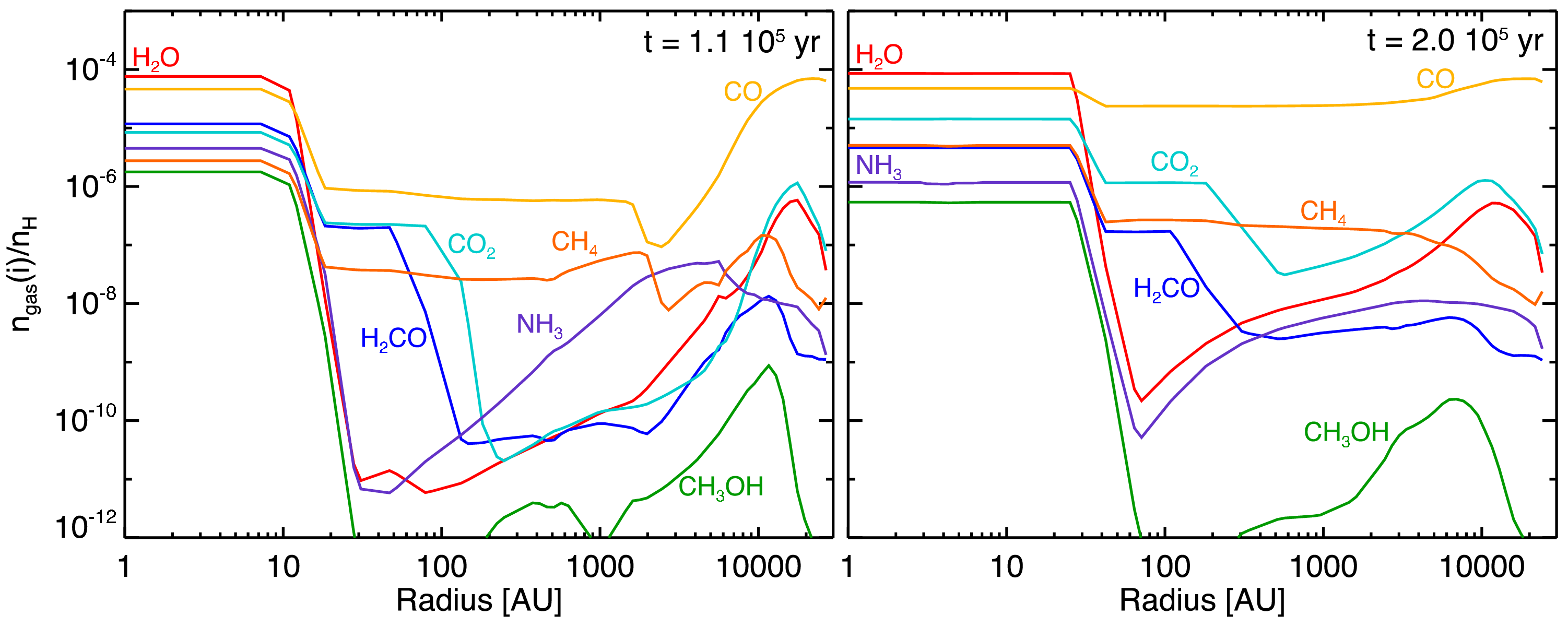}
\caption{Gas phase abundance profiles of the main ice components
   obtained for the collapsing core A-F at the beginning (left) and at the
   end (right) of the Class 0 phase.} 
\label{L1498class0_abu}
\end{figure*}

Once the free-fall timescale of the core is reached, the protostar
formation starts influencing the chemical abundance profiles through
the warm-up of the protostellar envelope. 
Figure \ref{L1498class0_abu} presents the gas phase abundance profiles
in the protostellar envelope of the collapsing core A-F for the main
ice components obtained at the beginning and at the end of the
Class 0 phase.  
As shown in section 3.1, the strong increase of the accretion rate
induces a quick increase of the temperature in the protostellar
envelope. 
The multilayer approach used to follow the evolution of interstellar ices
induces complex abundance profiles with double abundance jumps for
solid species that are more volatile than water 
\citep[see also][]{Vasyunin2013, Garrod2013}.

For species like CO, CO$_2$, CH$_4$, or H$_2$CO, the abundance
profiles can be distinguished into three regimes: \\  
i) the cold external envelope, where $T < T_{ev,i}$ (where $T_{ev,i}$
is the temperature of sublimation of species $i$), in which the
abundance profile in the gas phase of each species $i$ is governed by
the balance between its formation from gas phase reactions and
photo-evaporation and its destruction through the freeze-out on grains
and its photodissociation. 
The decrease of the abundances towards the protostar is due
to the increase of the density, as the accretion rate becomes more
efficient, and to the increase of the visual extinction, decreasing
the photodesorption rates. \\ 
ii) the cold intermediate envelope, where $T_{ev,i} < T < T_{ev,\textrm{H2O}}$,
in which a part of their icy content evaporates thermally when the
temperature reaches their temperature of sublimation through the
diffusion of volatile particles towards the surface.
However, most of their volatile content stays trapped into the ices. \\
iii) the warm inner envelope, where $T > T_{ev,\textrm{H2O}}$, in which all the
icy material is evaporated with water in the gas phase when $ T =
T_{ev,\textrm{H2O}} \sim 100 - 110$ K via the so-called
``volcano desorption''  \citep{Collings2004}. \\
The abundance profiles of species showing similar or higher binding
energies than water do not display the second regime.
Since the temperature in the protostellar envelope increases with the age of
the protostar while the density decreases, the gas phase abundances
tend to increase with time, inducing a disappearance of the double
desorption peaks for some species, and an increase of the size of the
hot corino region.   

\begin{figure*}[htp]
\centering 
\includegraphics[width=180mm]{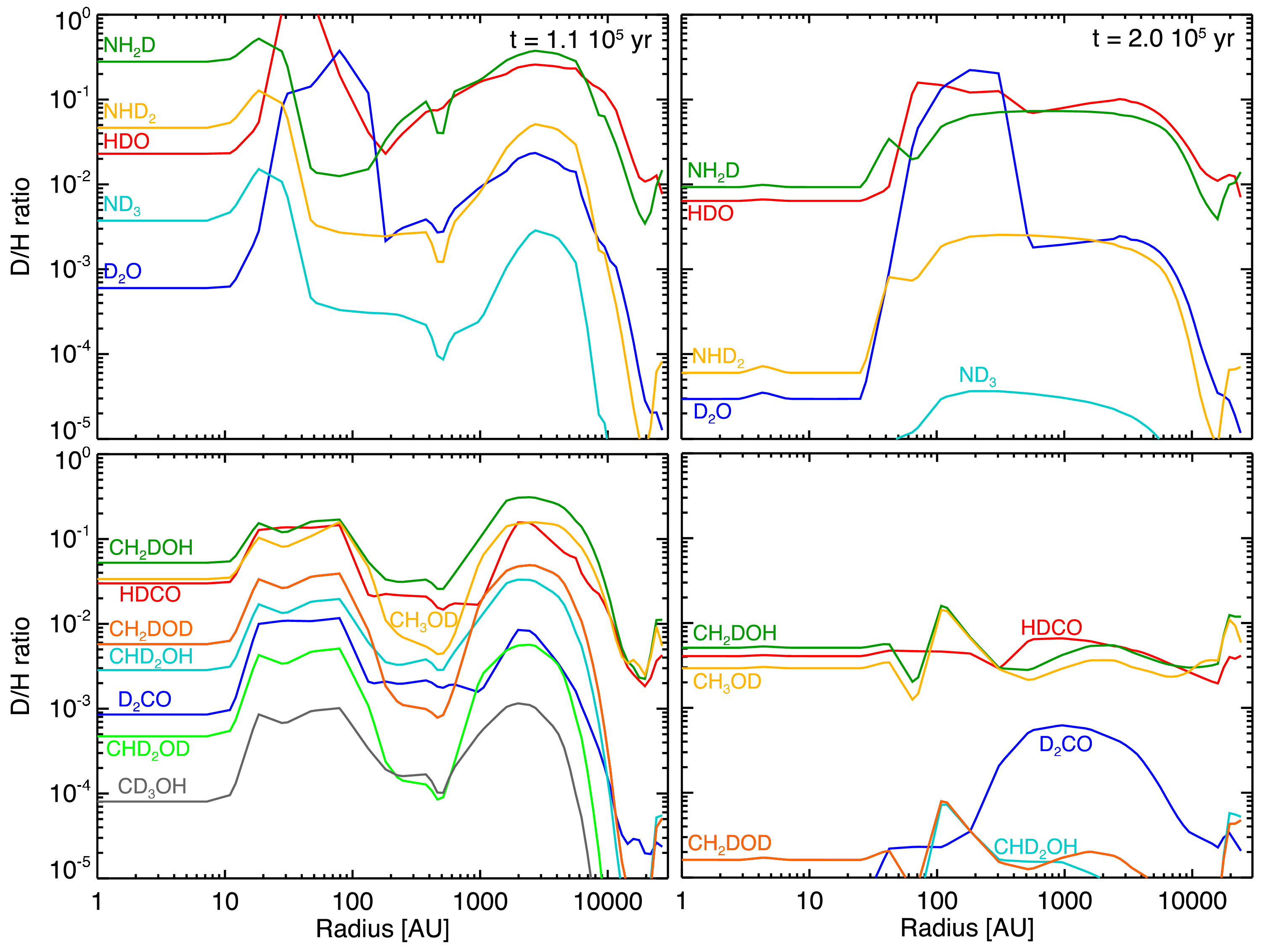}
\caption{Gas phase deuteration profiles of the main ice components
   obtained for the collapsing core A-F at the beginning and at the end of the Class 0 phase.}
\label{L1498class0_deut}
\end{figure*}

As it can be seen in Figure \ref{L1498class0_deut}, the deuteration of
all species but ammonia is higher in the external envelope than in the hot
corino. The D/H ratio predicted in the hot corino reflects the overall
deuteration of ices whereas the deuteration of the external
envelope is due to the sublimation of the highly deuterated ice
surface through non-thermal processes and eventually supported by a
gas phase chemistry.  
At the beginning of the Class 0 phase, the D/H ratio of simply
deuterated species is about 10 times higher in the cold envelope, at
$r = 3000$ AU, than in the hot corino, at $r = 10$ AU. 
This is in good agreement with the difference of deuteration between
icy and gaseous molecules predicted in dense cores. 
A peak of the deuteration is observed between 20 and 100 AU,
corresponding the zone where interstellar ices start to sublimate at
$T \sim 100$ K during the fast collapse. Since deuterated species are
mostly located at the surface, they mostly evaporate before their main
isotopologue which are rather located in the inner part of the ice,
increasing the D/H abundance ratio in the gas phase in the evaporation
zone.

During the evolution of the Class 0 phase, the deuteration of simply
and doubly deuterated species in the hot corino decrease by one
and more than two orders of magnitude respectively, due to the gradual
collapse of each shell of the envelope (see Table \ref{results_deut}
for deuteration values of various species). 
At the beginning of the Class 0 phase, the gas seen in the
hot corino ($1 < r < 30$ AU) comes from the shells located at $r \sim
5000$ AU whereas the warm gas modelled at the end of
the Class 0 phase originates from shells located at $r \sim 1.5 \times 10^4$
AU where the deuteration is lower.  
Although the deuteration of water in the hot corino decreases with the
age of the protostar, its deuteration remains roughly constant in
the external envelope. This is due to gas phase reactions and in
particular to the recombination reactions between highly deuterated OH
radicals, and their deuterated isotopologues, in the gas phase at $50
< T < 100$ K that form highly deuterated water.

\subsection{Consequences on complex organics deuteration}

\begin{figure*}[htp]
\centering 
\includegraphics[width=180mm]{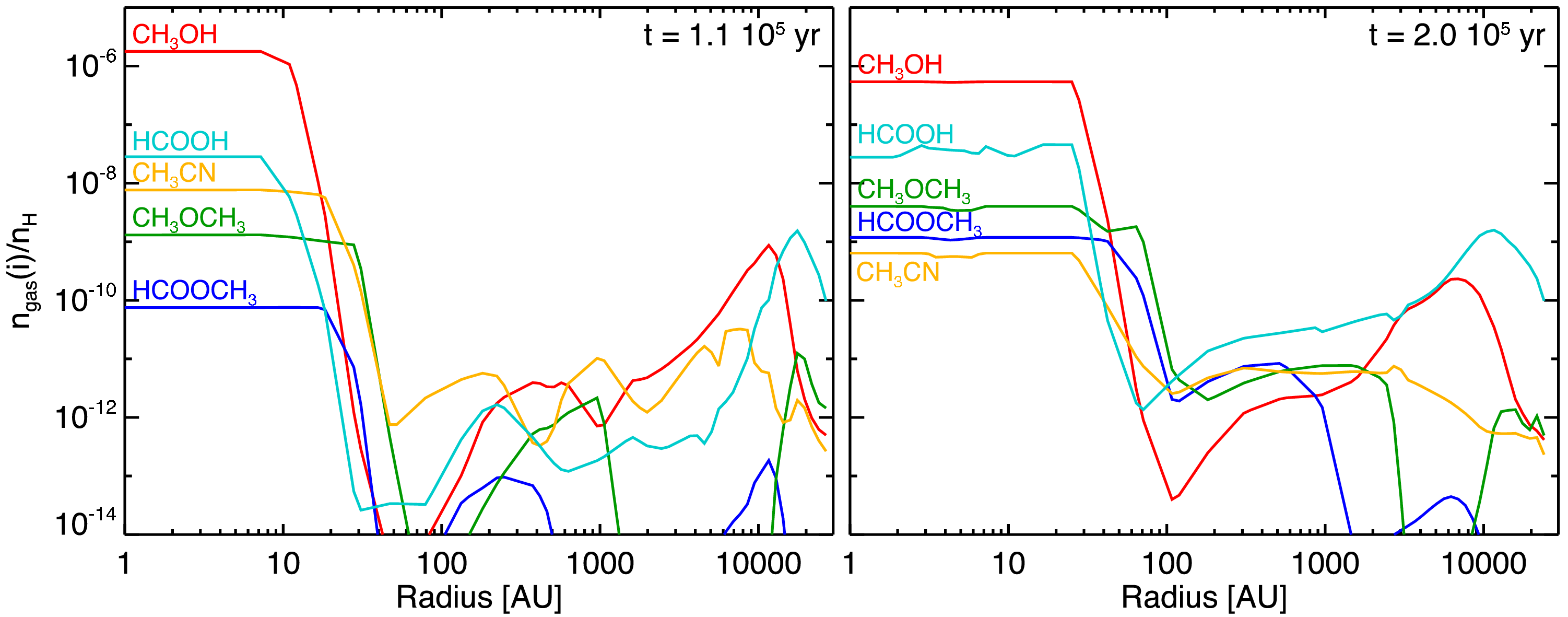}
\caption{Gas phase abundance profiles of selected COMs thought to be mainly
  formed at the surface of interstellar grains obtained for the
  collapsing core A-F at the beginning and at the end of the Class 0 phase.} 
\label{L1498_class0_COMs}
\end{figure*}

We modelled the formation and the deuteration of complex organics at
the surface of interstellar grains through radical recombinations and
hydrogenation reactions by incorporating the surface chemical network
of \citet{Garrod2008}. 
We only focused our study on a few prototype COMs; methyl formate
(HCOOCH$_3$), methyl cyanide (CH$_3$CN), di-methyl ether
(CH$_3$OCH$_3$), and formic acid (HCOOH) whose sub-millimetric
spectra of their deuterated isotopologues have recently been measured in
laboratories \citep{Margules2010, Cazzoli2011, Richard2013,
  Nguyen2013, Coudert2013}, allowing recent and future detections
towards low-mass protostars. 
Figure \ref{L1498_class0_COMs} presents the spatial distributions of
the gas phase abundances of methanol and the four COMs in the
protostellar envelope of the collapsing core A-F at the beginning
and the end of the Class 0 phase.

In our model, methyl formate and di-methyl ether are mostly formed
through radical recombination reactions (HCO + CH$_3$O for methyl
formate and CH$_3$ + CH$_3$O for di-methyl ether) at the surface of
warm interstellar grains ($30 < T < 50$ K), triggered by the
photodissociation of the main ice constituents from the cosmic ray
induced UV field. Formic acid is thought to be mostly formed in cold
ices in dark clouds through the reaction sequence
\begin{equation} 
\textrm{CO} + \textrm{OH} \rightarrow
\textrm{HO-CO}~ (+ \textrm{H})  \rightarrow \textrm{HCOOH} \label{reac_HCOOH} 
\end{equation}  
as suggested by laboratory and theoretical works \citep{Goumans2008,
  Ioppolo2011}. The reaction between HCO and OH also plays a minor
role.
Two main routes play a major role in the formation of methyl
cyanide as it can be formed either in cold ices through the
hydrogenation sequence of C$_2$N or during the warm-up phase through
the radical recombination between CH$_3$ and CN.

Complex organics desorb thermally in the hot corino when the
temperature reaches their temperature of sublimation. 
A part of the particles formed on ices also evaporates through
non-thermal processes at lower temperature, generating the complex
abundance profiles shown in Fig. \ref{L1498_class0_COMs}.  
The abundances of methyl formate and di-methyl ether tend to increase
with the age of the protostar because the timescale of the warm-up
phase (when $30 < T < 50$ K) required for an efficient formation of
COMs increases with the age of the protostar. 
Like water and carbon dioxide, the abundance of formic acid remains
roughly constant with the evolutionary stage because it is efficiently
formed throughout the precursor dark core, unlike methyl cyanide whose
abundance decreases by one order of magnitude because of its lower
formation at the edges of the core. 

The deuteration of the simply deuterated isotopologues of COMs in the
hot corino is shown in Table \ref{results_deut} along with the
deuteration of the main ice components. 
It can be seen that the deuteration roughly scales with the
deuteration of their parent molecules.  
Since methyl formate and di-methyl ether are mostly formed at the
external surface of warm ices from the photodissociation of highly
deuterated formaldehyde and methanol, their deuterium fractionation is
high (up to $\sim 20$ \%) at the beginning of the Class 0 phase and
decreases with time.
The fractionation of HCOOCH$_2$D is larger than that of DCOOCH$_3$
because formaldehyde, forming HCO (and DCO) through photodissociation,
is less deuterated than methanol, forming CH$_3$O (and its deuterated
isotopologues) at the surface of ices. 
A significant part of formic acid and methyl cyanide is formed in the less
deuterated phase of dark clouds, their D/H abundance ratio is
consequently lower (1 - 2 \% and 3.5 \%, respectively).
As for the main ice species, the deuteration of COMs decreases with
the age of the protostar from one to two orders of magnitude.

\section{Discussion}

\subsection{Comparison with previous models}

{One of the goals of this work is the study of the spatial and temporal
  evolutions of the deuterium chemistry within protostellar
  envelopes.
It is therefore worth comparing our results with those by \citet{Aikawa2012}
  although the two approaches differ slightly. 
 \citet{Aikawa2012} adopted a more simple astrochemical model,
  assuming a homogeneous ice chemistry and neglecting the spin states
  of H$_2$ and other ions important for the deuterium chemistry but
  they adopted a more sophisticated hydrodynamical model than ours.

The most striking difference is found for the spatial evolution of
abundances and deuterations within prestellar cores. 
Our model predicts stronger decreases of the abundances and
deuterations of the main icy species towards the edge of the core than
the model by Aikawa et al. (2012). 
These differences are likely due to the use of the multilayer approach
in this work which makes the ice chemistry much more dependent on the
physical parameters than the standard approach that assumes
homogeneous ices \citep[see][]{Taquet2012a}.
The temporal evolution of the gaseous deuteration predicted at the
center of protostellar envelopes is consequently much stronger in our
work than in the work by Aikawa et al. (2012).

The comparison of our ``early phase'' results (at $t=1.1 \times 10^5$
yr, see Table \ref{results_deut}) with those by Aikawa et al. (2012)
at $t=9 \times 10^4$ yr shows that the two works predict a similar
evolution of the water deuteration, with a D/H ratio of a few percents
in the inner region and $> 10$ \% in the cold external
envelope. 
However, the deuteration of ammonia, formaldehyde, and methanol is 2 to
20 times higher in our work, because of their late formation at the
surface of interstellar ices when the gas phase D/H ratio is high. 
We also note a discrepancy in the predictions of the deuteration of
the COMs since we predicted higher deuteration for methyl formate than for
formic acid and methyl cyanide ($> 10$ \% versus $1-4$ \%) while
\citet{Aikawa2012} predicted a higher deuteration for methyl cyanide
and formic acid (7 and 50 \%) than for methyl formate (1 \%). 
}

\subsection{Comparison with observations} 

\subsection{Abundances}

{Table \ref{params_ice} of the Appendix compares the chemical
composition of interstellar ices predicted by GRAINOBLE with infrared
observations of ices towards dense cores. Observational data is taken from
\citet{Oberg2011} who summarized results obtained by
\citet{Boogert2011} towards a sample of dense cores and compared them
with the {\it Spitzer} c2d Legacy ice survey carried out towards
low-mass protostars.  }
It can be seen that, even if surface parameters strongly influence the ice
composition as discussed in section \ref{influence_params}, the
standard model seems to be in fair agreement with the 
ice observations, since the predicted abundance of most of species are
close to the range of abundances given by the observations. 
However, the standard model tends to underpredict the formation of
CO$_2$ (10 \% versus 38 \%) in favour of CO (65 \% versus 31 \%), and
overestimates the H$_2$CO abundance compared to CH$_3$OH.  
An increase of the {external temperature} of 3 K ($T_{\textrm{ext}} = 17$ to 20
K) significantly increases the abundance of CO$_2$ and CH$_3$OH,
showing observable values.  
In all of the models, H$_2$CO is overproduced, implying that the
destruction of H$_2$CO, forming CH$_3$OH, is not efficient enough. 
Laboratory studies have shown that H and D atoms can penetrate into a
few surface layers \citep{Watanabe2004, Ioppolo2010}.
The incorporation of such diffusion processes between several reactive
surface layers in astrochemical models would, therefore, increase the
efficiency of hydrogenation reactions and decrease the abundance of
H$_2$CO in favour of CH$_3$OH.

Table \ref{results_COMs} compares the predicted abundances of selected
COMs with respect to that of methanol with the sub-millimetric
observations towards a sample of low-mass protostars performed
with single-dish telescopes. For a sake of consistency
between all the molecules, all  the observed values are derived from
rotational diagram analysis, giving column densities averaged over the
large beam of the single-dish telescope ($\theta_b = 10 - 30$ \arcsec)
that encompass the hot corino, and the cold external envelope.  
Absolute abundances are not presented since they strongly depend on
the choice of the H$_2$ column density, which is quite uncertain.
The predicted abundance ratios are, therefore, derived from
column densities averaged over a beam of a radius $R_b = 3000$ AU
($\sim 30$ \arcsec~ at 200 pc, the averaged distance of the observed
low-mass protostars).

\begin{table*}[htp]
\centering
\caption{Modelled and observed abundances (in \%) of COMs relative to CH$_3$OH in protostellar envelopes.
}
\begin{tabular}{l c c c c c c c}
\hline
\hline
	&	\multicolumn{2}{c}{Model ($R_b = 3000$ AU)}	&			\multicolumn{5}{c}{Observations}									\\
\hline															
	&	$t = 1.1 \times 10^5$ yr	&	$t = 2.0 \times 10^5$ yr	&	I16293	&	I2A	&	I4A	&	I4B	&	Ref.	\\
\hline															
H$_2$CO	&	870	&	1400	&	20 - 33	&	25 - 36	&	11 - 17	&	17 - 26	&	1, 2	\\
HCOOCH$_3$	&	0.011	&	0.44	&	21 - 57	&	$< 85$	&	30 - 91	&	2.6 - 31	&	3, 4, 5	\\
CH$_3$OCH$_3$	&	0.24	&	1.45	&	11 - 64	&
$< 52$	&	$< 21$	&	$< 19$	&	3, 4, 5	\\
 CH$_3$CN	&	1.7	&	0.59	&	0.50 - 1.50	&	0.68 - 1.6	&	0.9 - 1.8	&	0.98 - 1.9	&	3, 4, 5	\\
HCOOH	&	6.2	&	38	&	5.2 -6.5	&	$< 13$	&	0 - 12	&	$< 14$ 	&	3, 4, 5	\\
\hline
\end{tabular}
\label{results_COMs}
\tablecomments{
1: \citet{vanDishoeck1995} for I16293 
2: \citet{Maret2004} for I2A, I4A, I4B
3: \citet{Cazaux2003} for I16293
4: \citet{Bottinelli2004} for I4A
5: \citet{Bottinelli2007} for I2A and I4B
} 
\end{table*}

As for ices, our model strongly overpredicts the abundance of H$_2$CO
in protostellar envelopes.
We predict a gaseous H$_2$CO/CH$_3$OH abundance ratio of $\sim 10$
while the gaseous H$_2$CO/CH$_3$OH abundance ratio observed in the gas 
phase of protostellar envelopes is consistent with that observed in ices
($\sim 30$ \%), suggesting that the gas phase chemistry does not alter
significantly their relative abundance after their evaporation.
The comparison between predicted and observed abundances of COMs
gives contrasting results. On the one hand, the abundance ratios of
CH$_3$CN and HCOOH are in good agreement with observations. 
While the predicted CH$_3$CN abundance ratio tends to
decrease with the age of the protostar, it remains within the range of
observed values, which are roughly constant from source to source
(between 0.5 and 1.9 \%). 
The HCOOH abundance ratio predicted at the beginning of the Class 0
stage lie in the range of observed values while the late prediction
slightly overestimates the ratio.
On the other hand, the abundance of HCOOCH$_3$ and CH$_3$OCH$_3$,
considered as late-stage molecules, is underpredicted by our model by
one to three orders of magnitude, showing that the formation of these
molecules is still not completely understood. The predicted abundances
of these two COMs barely exceed 1 \% with respect to methanol whereas
observations show that their abundance should easily exceed 10 \% in
some cases.  
Because of the fast collapse considered in our one-dimension model,
molecules that are evaporated at $T > 100$ K stay in the hot corino
for $\sim 100$ yr before being accreted by the protostar. Warm gas
phase chemistry is not efficient enough to produce complex organics in
significant quantities \citep{Horn2004}.
Experimental studies and molecular dynamics simulations have
demonstrated that UV photons can penetrate into the inner part of
interstellar ices, up to $\sim 100$ monolayers \citep{Gerakines2001,
  Andersson2008}, triggering a bulk chemistry induced by the mobility
of photoproducts. \citet{Garrod2013} introduced a three-phase approach
in order to treat the bulk chemistry, allowing the formation of
complex organics in the ice mantle. However, the predicted abundance
ratios of HCOOCH$_3$ and CH$_3$OCH$_3$ do not exceed 1 \% for the
three types of physical models considered in his work, suggesting that
bulk chemistry does not seem to be efficient enough to produce complex
organics.

\subsubsection{Deuterium fractionations}

The deuterium fractionation of various species predicted by our model
is compared with sub-millimetric observations of low-mass protostars
in Table \ref{results_deut}. As for Table \ref{results_COMs}, the
predicted deuterations are derived from the beam-averaged column
densities for two typical sizes. $R_b = 50$ AU corresponds to a small
beam of 0.4 \arcsec~obtained with sub-millimetric interferometers,
such as PdBi, SMA, or ALMA, while $R_b = 3000$ AU corresponds to a large
single-dish beam of 30 \arcsec. 
As already mentioned in the introduction, the deuteration of water has
been extensively studied in a sample of low-mass protostars with
different telescopes (ground-based and space telescopes, single-dish
and interferometers) and with different methods (standard LTE and
non-LTE LVG models based on constant and spherical physical
structures) giving contrasting results. 
By taking all the results together into account, the warm
[HDO]/[H$_2$O] ratio originating from the hot corino shows a large
range of values, from 0.04 to 8 \%, while the water deuteration of the
cold envelope is about 0.3 - 18 \% \citep[see the discussion
in][]{Coutens2013b} which are in good agreement with our predictions
since the predicted HDO/H$_2$O and D$_2$O/H$_2$O abundance ratios
decrease with the age of the protostar within the large range of
values given by the observations. 
However, the more recent observations using sub-mm interferometers and
the {\it Herschel Space Observatory} have refined the results.
Interferometric observations, with high angular resolution
allowing astronomers to probe the hot corino region of protostellar
envelopes, have yielded low HDO/H$_2$O abundance ratios of $0.05 -
0.3$ \% \citep{Persson2013, Persson2014} while {\it Herschel} observations,
more sensitive to the external envelope, show that the cold water
shows a higher deuteration of $\sim 1$ \% \citep{Coutens2013b}. 
This trend is confirmed by our astrochemical model. The deuteration in
the hot corino scales with the overall deuteration of interstellar
ices and includes the poorly deuterated inner mantle whereas the 
deuteration of the cold envelope, induced by the evaporation of highly
deuterated ice surfaces and supported by the cold gas phase chemistry,
remains roughly constant and shows higher values of a few \%.
The deuteration of water, therefore, increases with the size of the beam, by a factor
2 to 10, showing that the highly deuterated external and cold
protostellar envelopes can contribute significantly to the overall deuteration
observed with single-dish telescopes, in spite of their low abundance. 

\begin{table*}[htp]
\centering
\caption{Modelled and observed deuterium fractionations (in \%) of selected species in
  protostellar envelopes and comets.
}
\begin{tabular}{l c c c c c c c c c c}
\hline
\hline
	&	\multicolumn{4}{c}{Model}							&	\multicolumn{3}{c}{Low-mass protostars}					&	\multicolumn{3}{c}{Comets}					\\
\hline																					
	&	\multicolumn{2}{c}{$t = 1.1 \times 10^5$ yr}			&	\multicolumn{2}{c}{$t = 2.0 \times 10^5$ yr}			&	D/H 	&	Source	&	Ref.	&	D/H 	&	Source	&	Ref.	\\
	&	$50$ AU	&	$3000$ AU	&	$50$ AU	&	$3000$ AU	&		&		&		&		&		&		\\
\hline																					
HDO	&	2.3	&	4.5	&	0.65	&	2.2	&	In: 0.04 - 8	&	4 protostars	&	1 - 8	&	0.032 - 0.092	&	8 comets	&	22, 23	\\
	&		&		&		&		&	Out: 0.3 - 18	&		&	3, 6, 9	&		&		&		\\
D$_2$O	&	0.062	&	0.11	&	0.0036	&	0.043	&	In: 0.007	&	IRAS 16293	&	3	&	/	&	/	&	/	\\
	&		&		&		&		&	Out: 0.03	&		&	3	&		&		&		\\
HDS	&	4	&	3.8	&	0.54	&	0.65	&	5 - 15	&	IRAS 16293	&	10	&	$< 40$	&	Hale-Bopp	&	24	\\
D$_2$S	&	0.28	&	0.28	&	0.00094	&	0.0021	&	0.6 - 1.8	&	IRAS 4A	&	11	&	/	&	/	&	/	\\
NH$_2$D	&	29	&	31	&	0.99	&	4.3	&	7.0 - 10	&	2 protostars	&	10, 12	&	$< 24$	&	Hale-Bopp	&	24	\\
NHD$_2$	&	5.1	&	3.8	&	0.0074	&	0.11	&	2.6 - 3.0	&	IRAS 16293	&	13	&	/	&	/	&	/	\\
ND$_3$	&	0.44	&	0.21	&	0.000057	&	0.0012	&	0.09	&	NGC 1333	&	14	&	/	&	/	&	/	\\
DCN	&	18	&	18	&	0.46	&	1.1	&	0.40 - 7.1	&	17 protostars	&	15, 16	&	0.2	&	Hale-Bopp	&	25	\\
HDCO	&	4.1	&	3.8	&	0.42	&	0.43	&	2.2 - 170	&	7 protostars	&	15, 17	&	$< 10$	&	Hale-Bopp	&	24	\\
D$_2$CO	&	0.18	&	0.16	&	0.00064	&	0.0092	&	4.6 - 44	&	7 protostars	&	17	&	/	&	/	&	/	\\
CH$_2$DOH	&	5.5	&	5.6	&	0.51	&	0.51	&	37 - 65	&	4 protostars	&	17	&	$< 3.2$	&	Hale-Bopp	&	24	\\
CH$_3$OD	&	3.5	&	3.5	&	0.3	&	0.3	&	1.6 - 4.7	&	4 protostars	&	17	&	$< 12$	&	Hale-Bopp	&	24	\\
CHD$_2$OH	&	0.31	&	0.31	&	0.00078	&	0.00076	&	7.4 - 25	&	3 protostars	&	17	&	/	&	/	&	/	\\
CD$_3$OH	&	0.0092	&	0.0092	&	0.0016	&	0.0016	&	0.20 - 1.4	&	IRAS 16293	&	18	&	/	&	/	&	/	\\
DCOOCH$_3$	&	9.2	&	/	&	0.0025	&	/	&	15	&	IRAS 16293	&	19	&	/	&	/	&	/	\\
HCOOCH$_2$D	&	21	&	/	&	0.65	&	/	&	/	&	/	&	/	&	/	&	/	&	/	\\
CH$_2$DOCH$_3$	&	22	&	/	&	0.61	&	/	&	15	&	IRAS 16293	&	20	&	/	&	/	&	/	\\
CH$_2$DCN	&	3.6	&	/	&	0.15	&	/	&	1.3	&	IRAS 16293	&	21	&	/	&	/	&	/	\\
DCOOH	&	2.3	&	/	&	0.58	&	/	&	/	&	/	&	/	&	/	&	/	&	/	\\
HCOOD	&	1	&	/	&	0.66	&	/	&	/	&	/	&	/	&	/	&	/	&	/	\\
\hline
\end{tabular}
\label{results_deut}
\tablecomments{
1: \citet{Parise2005}
2: \citet{Coutens2012}
3: \citet{Coutens2013a}
4: \citet{Persson2013}
5: \citet{Visser2013}
6: \citet{Taquet2013b}
7: \citet{Persson2014}
8: \citet{Coutens2013b}
9: \citet{Liu2011}
10: \citet{vanDishoeck1995}
11: \citet{Vastel2003}
12: \citet{Sakai2009}
13: \citet{Loinard2001}
14: \citet{vanderTak2002}
15: \citet{Roberts2002}
16: \citet{Jorgensen2004}
17: \citet{Parise2006}
18: \citet{Parise2004}
19: \citet{Demyk2010}
20: \citet{Richard2013}
21: C. Kahane (private communication), taken from the analysis of the TIMASSS survey \citet{Caux2011}
22: \citet{Weaver2008}
23: \citet{Hartogh2011}
24: \citet{Crovisier2004}
25: \citet{Meier1998}
} 
\end{table*}

Formaldehyde and methanol show very high deuterium fractionations,
with observed HDCO/H$_2$CO and CH$_2$DOH/CH$_3$OH abundance ratios of
$\sim 20$ and $\sim 50$ \%, respectively. 
Such high fractionations are not reproduced by our model, since the
highest D/H ratios of formaldehyde and methanol, reached at
the beginning of the Class 0 phase, are $\sim 4$ and $\sim 5.5$ \%,
respectively, a factor of 4 and 9 lower than the observed values. 
For all the protostars observed by \citet{Parise2006}, the rotational
diagram analysis give lower rotational temperatures for deuterated
formaldehyde and methanol isotopologues than for their main
isotopologue, suggesting that deuterated species would originate from
colder regions than their main isotopologue. 
As shown in Fig. \ref{L1498core_deut}, the high observed deuteration
can indeed be reached in the external envelope, between $r = 1000$ and
5000 AU but the absolute abundances of deuterated formaldehyde and
methanol in the external envelope are too low to increase the
beam-averaged deuteration. 
The CH$_3$OD/CH$_3$OH abundance ratio predicted at the beginning of
the Class 0 phase lies in the range of observational values. 
CH$_3$OD is thought to be formed only via addition reactions since
abstraction reactions that enhance the deuteration of methanol seem to
occur only on the methyl group of methanol. 
The abstraction reactions are, therefore, not efficient enough to increase
the abundances of CH$_2$DOH, CHD$_2$OH, and CD$_3$OH. 
As shown in Fig. \ref{poplay_layers}, methanol is present in a
significant part of ices, and not only in the outermost layers, as it
was predicted in the pseudo-time dependent calculations assuming
constant physical conditions \citep{Taquet2012b}. 
While the multilayer approach considered here only considers one reactive
surface, the diffusion of H and D atoms into the bulk would tend to increase the
deuteration of methanol and the [CH$_2$DOH]/[CH$_3$OD] abundance ratio
since more CH$_3$OH molecules will be available to react. 
Other processes, such as H/D exchange between water and methanol, or
gas phase reactions can also play an important role in the evolution of the
[CH$_2$DOH]/[CH$_3$OD] ratio \citep{Charnley1997, Ratajczak2009}.

The deuteration of methyl formate, methyl cyanide, and dimethyl ether,
observed towards I-16293, are in fair agreement with our model
predictions. 
They confirm that methyl formate and dimethyl ether, thought to be
mainly formed from highly deuterated ices during the warm-up phase
in protostellar envelopes, show higher deuterations than methyl cyanide
which would also be formed in the cold anterior phase.
This comparison tends to valid the formation mechanisms and the moment
of formation of these three molecules. 
Now that sub-millimetric spectra have been measured and constrained by
laboratory experiments for other COMs, the deuterium fractionation of
COMs observed in protostellar envelopes can be used as a chemical
tracer to understand their chemical history through the
interpretation of their moment and their mechanisms of formation. 


\section{Conclusions}

In this work, we have theoretically followed the chemical evolution of
interstellar ices from the translucent phases of molecular clouds to
the envelopes of low-mass protostars by focusing on their deuterium
fractionation. For this purpose, we coupled a gas-grain astrochemical
model, following the multilayer formation and the evaporation of ices,
with a one-dimensional dynamical/radiative transfer model of
collapsing core. 
The main conclusions of this work are summarized below.

1) Interstellar ices predicted by our model present a chemical
heterogeneity, in agreement with infrared observations of ices. Water and
carbon dioxide are mostly formed in the inner part of the ice mantle
while carbon monoxide, formaldehyde, and methanol are closer to the ice
surface. Ammonia is present in the innermost and outermost ice layers.

2) Interstellar ices display an evolution of the deuteration within
the mantle, the deuteration of all species increasing towards the
surface by several orders of magnitude. 
This is due to the slow formation of ices from the translucent phases
of molecular clouds to the dense and cold phase of dense cores, and
the slow decrease of two chemical parameters, the gas phase
abundance of CO and the ortho/para ratio of H$_2$, that limit the
deuterium chemistry,  

3) The multilayer approach used to follow the formation and the
evaporation of ices induces double abundance jumps of icy species more
volatile than water in protostellar envelopes. A part of the volatile
content stays trapped into the ices when the temperature exceeds the
temperature of sublimation and is released when water ice evaporates
at $T \sim 100$ K.  

4) The deuteration of most hydrogenated species evolves within the
envelopes of low-mass protostars. The deuteration predicted in the hot
corino is generally lower than the deuteration of the external
envelope. The difference of deuteration in the warm and cold envelopes
is explained by the gradient of deuteration within interstellar
ices. Only the external ice layers evaporate in the cold envelope
through non-thermal processes while the inner part of ice mantles
evaporates only in the hot corino.

5) The deuteration tends to decrease with the evolutionary stage of
the protostar, due to the gradual collapse of the external shells of the
protostellar envelope that are less deuterated because of their
lower density. 

6) We have modelled the deuteration of some complex organics (COMs)
that are believed to be mainly formed at the surface of interstellar
grains. The deuteration of COMs is very sensitive to their formation
pathways and to their moment of formation, Consequently, the deuterium
fractionation can be used as a tracer to interpret their chemical history.

7) The comparison with the observations gives contrasting results. The
model successfully reproduces the evolution of the D/H abundance ratio
of water  within protostellar envelopes as recently observed with
ground-based interferometers and the {\it Herschel Space
  Observatory}. It also reproduces the different D/H ratio of COMs
recently observed with single-dish telescopes. 
However, it underestimates the very high deuteration of formaldehyde
and methanol, suggesting that the understanding and the modelling of
interstellar ices is still not well understood. 

\begin{acknowledgements}
The kinetic data we used have been downloaded from the online database
KIDA (Wakelam et al. 2012, http://kida.obs.u-bordeaux1.fr). 
This work was supported by NASA's Origins of Solar Systems and
Exobiology Programs.
V. T. acknowledges support from the NASA postdoctoral program. 
O.S. acknowledges support from the Academy of Finland grant 250741.

\end{acknowledgements}

 \newpage
\appendix

\begin{table*}[htp]
\centering
\caption{Observed and predicted abundances (in \%) of solid species with respect to
  water ice at the end of the prestellar core stage.
}
\begin{tabular}{l c c c c c c c}
\hline
\hline
Model	&	H$_2$O	&	CO$_2$	&	CO	&	H$_2$CO +	&	CH$_3$OH	&	NH$_3$	&	CH$_4$	\\
	&		&		&		&	 HCOOH	&		&		&		\\
\hline															
A	&	100	&	10.5	&	64.8	&	16.3	&	2.4	&	6.0	&	3.0	\\
B	&	100	&	10.9	&	93.9	&	16.3	&	1.8	&	12.0	&	1.5	\\
\hline															
A, $n_{\textrm{H,ini}} = 3 \times 10^3$ cm$^{-3}$	&	100	&	10.5	&	64.8	&	16.3	&	2.4	&	6.0	&	3.0	\\
A, $n_{\textrm{H,ini}} = 1 \times 10^4$ cm$^{-3}$	&	100	&	7.7	&	56.7	&	13.1	&	1.9	&	13.2	&	6.7	\\
\hline															
A-F, $A_{\textrm{V,ext}} = 1$ mag	&	100	&	8.6	&	74.9	&	12.5	&	1.3	&	5.2	&	1.0	\\
A-F, $A_\textrm{V,ext} = 2$ mag 	&	100	&	10.5	&	64.8	&	16.3	&	2.4	&	6.0	&	3.0	\\
\hline															
A, $T_{\textrm{ext}}$ = 14 K	&	100	&	7.5	&	81.7	&	11.3	&	1.0	&	6.0	&	0.5	\\
A, $T_{\textrm{ext}}$ = 17 K	&	100	&	10.5	&	64.8	&	16.3	&	2.4	&	6.0	&	3.0	\\
A, $T_{\textrm{ext}}$ = 20 K	&	100	&	26.2	&	33.7	&	19.6	&	10.9	&	6.0	&	7.9	\\
\hline															
A, $t_{\textrm{BC}} = 1.7 \times 10^6$ yr 	&	100	&	10.5	&	64.8	&	16.3	&	2.4	&	6.0	&	3.0	\\
A, $t_{\textrm{BC}} = 3.4 \times 10^6$ yr 	&	100	&	8.7	&	45.9	&	13.3	&	2.3	&	10.0	&	5.9	\\
\hline															
A-F, $E_d/E_b = 0.5$ 	&	100	&	14.9	&	5.4	&	7.8	&	20.8	&	4.5	&	20.0	\\ 
A-F, $E_d/E_b = 0.65$	&	100	&	10.5	&	64.8	&	16.3	&	2.4	&	6.0	&	3.0	\\
A-F, $E_d/E_b = 0.8$ 	&	100	&	9.6	&	93.3	&	1.4	&	0.0	&	6.3	&	0.4	\\
\hline															
A-F, $E_b = 400$ K 	&	100	&	12.1	&	37.4	&	19.7	&	12.1	&	4.4	&	6.6	\\
A-F, $E_b = 500$ K 	&	100	&	10.5	&	64.8	&	16.3	&	2.4	&	6.0	&	3.0	\\
A-F, $E_b = 600$ K	&	100	&	34.3	&	73.7	&	13.3	&	1.5	&	8.4	&	0.8	\\
\hline															
A-F, o/p(H$_2$)$_{\textrm{ini}}$ = 10$^{-4}$	&	100	&	10.5	&	64.7	&	16.2	&	2.4	&	6.0	&	3.0	\\
A-F, o/p(H$_2$)$_{\textrm{ini}}$ = 10$^{-2}$	&	100	&	10.5	&	64.8	&	16.3	&	2.4	&	6.0	&	3.0	\\
A-F, o/p(H$_2$)$_{\textrm{ini}}$ = 3	&	100	&	10.5	&	64.9	&	16.4	&	2.4	&	5.8	&	3.0	\\
\hline															
Observations 	&	100	&	$38_{32}^{41}$	&	$31$	&	$2.8_{2.4}^{3.3}$	&	$8_7^{10}$	&	$5_4^6$	&	$5_4^7$	\\
\hline															
\end{tabular}
\label{params_ice}
\end{table*}

\begin{table*}[htp]
\centering
\caption{Observed and predicted deuterations in the gas phase and in
  ices at the end of the prestellar core stage.
}
\begin{tabular}{l c c c c c c c c}
\hline
\hline
	&	\multicolumn{8}{c}{Gas phase deuteration}
        \\
\hline
Model	&	p-D$_2$H$^+$/-oH$_2$D$^+$	&	N$_2$D$^+$	&	HDO	&	D$_2$O	&	HDCO	&	D$_2$CO	&	CH$_2$DOH	&	CHD$_2$OH	\\
\hline																	
A	&	1.60(-01)	&	7.76(-01)	&	2.40(-01)	&	1.80(-02)	&	1.02(-01)	&	3.32(-03)	&	2.30(-01)	&	1.60(-02)	\\
B	&	1.73(-01)	&	1.75(+00)	&	2.58(-01)	&	4.57(-02)	&	1.11(-01)	&	1.28(-03)	&	9.00(-02)	&	2.56(-03)	\\
\hline																	
A, $n_{\textrm{H,ini}} = 3 \times 10^3$ cm$^{-3}$	&	1.60(-01)	&	7.76(-01)	&	2.40(-01)	&	1.80(-02)	&	1.02(-01)	&	3.32(-03)	&	2.30(-01)	&	1.60(-02)	\\
A, $n_{\textrm{H,ini}} = 1 \times 10^4$ cm$^{-3}$	&	2.78(-01)	&	1.28(+00)	&	2.46(-01)	&	1.97(-02)	&	9.51(-02)	&	4.09(-03)	&	3.00(-01)	&	3.01(-02)	\\
\hline																	
A-F, $A_{\textrm{V,ext}} = 1$ mag	&	1.42(-01)	&	6.86(-01)	&	2.37(-01)	&	1.71(-02)	&	9.58(-02)	&	2.84(-03)	&	2.03(-01)	&	1.22(-02)	\\
A-F, $A_{\textrm{V,ext}} = 2$ mag	&	1.60(-01)	&	7.76(-01)	&	2.40(-01)	&	1.80(-02)	&	1.02(-01)	&	3.32(-03)	&	2.30(-01)	&	1.60(-02)	\\
\hline																	
A, $T_{\textrm{ext}} =$ 14 K	&	1.16(-01)	&	6.98(-01)	&	2.15(-01)	&	1.61(-02)	&	4.00(-02)	&	9.60(-04)	&	7.38(-02)	&	1.44(-03)	\\
A, $T_{\textrm{ext}} =$ 17 K	&	1.60(-01)	&	7.76(-01)	&	2.40(-01)	&	1.80(-02)	&	1.02(-01)	&	3.32(-03)	&	2.30(-01)	&	1.60(-02)	\\
A, $T_{\textrm{ext}} =$ 20 K	&	2.27(-01)	&	7.47(-01)	&	2.49(-01)	&	1.70(-02)	&	1.74(-01)	&	8.54(-03)	&	3.49(-01)	&	4.12(-02)	\\
\hline																	
A, $t_{\textrm{BC}} = 1.7 \times 10^6$ yr 	&	1.60(-01)	&	7.76(-01)	&	2.40(-01)	&	1.80(-02)	&	1.02(-01)	&	3.32(-03)	&	2.30(-01)	&	1.60(-02)	\\
A, $t_{\textrm{BC}} = 3.4 \times 10^6$ yr 	&	3.11(-01)	&	1.18(+00)	&	2.33(-01)	&	1.09(-02)	&	8.65(-02)	&	3.84(-03)	&	3.43(-01)	&	3.93(-02)	\\
\hline																	
A-F, $E_d/E_b = 0.5$ 	&	1.13(-01)	&	2.94(-01)	&	2.17(-01)	&	8.63(-03)	&	1.14(-01)	&	7.87(-03)	&	5.36(-01)	&	1.33(-01)	\\
A-F, $E_d/E_b = 0.65$	&	1.60(-01)	&	7.76(-01)	&	2.40(-01)	&	1.80(-02)	&	1.02(-01)	&	3.32(-03)	&	2.30(-01)	&	1.60(-02)	\\
A-F, $E_d/E_b = 0.8$ 	&	1.81(-01)	&	1.10(+00)	&	2.33(-01)	&	1.88(-02)	&	1.86(-02)	&	8.31(-04)	&	4.64(-02)	&	5.24(-04)	\\
\hline																	
A-F, $E_b = 400$ K 	&	2.01(-01)	&	8.89(-01)	&	2.43(-01)	&	1.45(-02)	&	1.32(-01)	&	5.06(-03)	&	2.68(-01)	&	2.30(-02)	\\
A-F, $E_b = 500$ K 	&	1.60(-01)	&	7.76(-01)	&	2.40(-01)	&	1.80(-02)	&	1.02(-01)	&	3.32(-03)	&	2.30(-01)	&	1.60(-02)	\\
A-F, $E_b = 600$ K	&	1.19(-01)	&	4.56(-01)	&	2.20(-01)	&	1.49(-02)	&	9.11(-02)	&	3.53(-03)	&	2.69(-01)	&	2.11(-02)	\\
\hline																	
A-F, o/p(H$_2$)$_{\textrm{ini}}$ = 10$^{-4}$	&	1.67(-01)	&	7.09(-01)	&	2.38(-01)	&	1.76(-02)	&	9.70(-02)	&	3.08(-03)	&	2.30(-01)	&	1.62(-02)	\\
A-F, o/p(H$_2$)$_{\textrm{ini}}$ = 10$^{-2}$	&	1.60(-01)	&	7.76(-01)	&	2.40(-01)	&	1.80(-02)	&	1.02(-01)	&	3.32(-03)	&	2.30(-01)	&	1.60(-02)	\\
A-F, o/p(H$_2$)$_{\textrm{ini}}$ = 3	&	1.54(-02)	&	4.24(-02)	&	1.06(-01)	&	6.89(-04)	&	1.29(-02)	&	6.41(-04)	&	9.09(-03)	&	2.43(-05)	\\
\hline																	
Observations L1498	&	/	&	4.00(+00)	&		&		&		&		&		&		\\
\hline																	
Observations L1544	&	$< 0.80$	&	2.30(-01)	&		&		&	3.20(-01)	&	4.00(-02)	&		&		\\
\hline																	
\hline																	
	&	\multicolumn{8}{c}{Ice deuteration}
        \\	
\hline
	&		&		&	HDO	&	D$_2$O	&	HDCO	&	D$_2$CO	&	CH$_2$DOH	&	CHD$_2$OH	\\
\hline																	
A	&		&		&	2.41(-02)	&	5.24(-04)	&	2.54(-02)	&	4.26(-04)	&	3.92(-02)	&	1.23(-03)	\\
B	&		&		&	2.14(-02)	&	2.59(-04)	&	1.57(-02)	&	9.59(-05)	&	2.29(-02)	&	1.55(-02)	\\
\hline																	
A, $n_{\textrm{H,ini}} = 3 \times 10^3$ cm$^{-3}$	&		&		&	2.41(-02)	&	5.24(-04)	&	2.54(-02)	&	4.26(-04)	&	3.92(-02)	&	1.23(-03)	\\
A, $n_{\textrm{H,ini}} = 1 \times 10^4$ cm$^{-3}$	&		&		&	2.10(-02)	&	5.96(-04)	&	2.83(-02)	&	7.20(-04)	&	6.07(-02)	&	3.19(-03)	\\
\hline																	
A-F, $A_{\textrm{V,ext}} = 1$ mag	&		&		&	2.69(-02)	&	5.17(-04)	&	3.41(-02)	&	5.00(-04)	&	6.12(-02)	&	1.80(-03)	\\
A-F, $A_{\textrm{V,ext}} = 2$ mag	&		&		&	2.41(-02)	&	5.24(-04)	&	2.54(-02)	&	4.26(-04)	&	3.92(-02)	&	1.23(-03)	\\
\hline																	
A, $T_{\textrm{ext}} =$ 14 K	&		&		&	2.47(-02)	&	2.75(-04)	&	2.20(-02)	&	1.24(-04)	&	3.42(-02)	&	3.70(-04)	\\
A, $T_{\textrm{ext}} =$ 17 K	&		&		&	2.41(-02)	&	5.24(-04)	&	2.54(-02)	&	4.26(-04)	&	3.92(-02)	&	1.23(-03)	\\
A, $T_{\textrm{ext}} =$ 20 K	&		&		&	2.36(-02)	&	6.91(-04)	&	1.78(-02)	&	5.31(-04)	&	1.42(-02)	&	6.52(-04)	\\
\hline																	
A, $t_{\textrm{BC}} = 1.7 \times 10^6$ yr	&		&		&	2.41(-02)	&	5.24(-04)	&	2.54(-02)	&	4.26(-04)	&	3.92(-02)	&	1.23(-03)	\\
A, $t_{\textrm{BC}} = 3.4 \times 10^6$ yr	&		&		&	2.32(-02)	&	6.62(-04)	&	3.35(-02)	&	7.60(-04)	&	6.01(-02)	&	2.69(-03)	\\
\hline																	
A-F, $E_d/E_b = 0.5$	&		&		&	1.98(-02)	&	1.53(-04)	&	1.46(-02)	&	2.48(-04)	&	3.15(-02)	&	2.12(-03)	\\
A-F, $E_d/E_b = 0.65$	&		&		&	2.41(-02)	&	5.24(-04)	&	2.54(-02)	&	4.26(-04)	&	3.92(-02)	&	1.23(-03)	\\
A-F, $E_d/E_b = 0.8$	&		&		&	1.45(-02)	&	9.67(-05)	&	7.38(-03)	&	2.02(-05)	&	6.78(-03)	&	3.33(-05)	\\
\hline																	
A-F, $E_b = 400$ K	&		&		&	1.97(-02)	&	5.10(-04)	&	1.53(-02)	&	3.76(-04)	&	1.04(-02)	&	3.52(-04)	\\
A-F, $E_b = 500$ K	&		&		&	2.41(-02)	&	5.24(-04)	&	2.54(-02)	&	4.26(-04)	&	3.92(-02)	&	1.23(-03)	\\
A-F, $E_b = 600$ K	&		&		&	2.86(-02)	&	6.16(-04)	&	3.59(-02)	&	7.04(-04)	&	6.38(-02)	&	2.37(-03)	\\
\hline																	
A-F, o/p(H$_2$)$_{\textrm{ini}}$ = 10$^{-4}$	&		&		&	3.62(-02)	&	9.32(-04)	&	3.86(-02)	&	6.86(-04)	&	5.62(-02)	&	1.89(-03)	\\
A-F, o/p(H$_2$)$_{\textrm{ini}}$ = 10$^{-2}$	&		&		&	2.41(-02)	&	5.24(-04)	&	2.54(-02)	&	4.26(-04)	&	3.92(-02)	&	1.23(-03)	\\
A-F, o/p(H$_2$)$_{\textrm{ini}}$ = 3	&		&		&	1.33(-03)	&	1.59(-06)	&	6.35(-04)	&	5.43(-07)	&	1.06(-03)	&	1.35(-06)	\\
\hline																	
\end{tabular}
\label{params_deut}
\end{table*}																							


\end{document}